\documentclass[iop]{emulateapj}

\usepackage{textcomp}
\newcommand{\m}{\,\hbox{m}}
\newcommand{\mm}{\,\hbox{mm}}
\newcommand{\km}{\,\hbox{km}}
\newcommand{\mum}{\,\hbox{\textmu{}m}}
\newcommand{\cm}{\,\hbox{cm}}

\newcommand{\AU}{\,\hbox{AU}}
\newcommand{\g}{\,\hbox{g}}
\newcommand{\s}{\,\hbox{s}}
\newcommand{\pc}{\,\hbox{pc}}
\newcommand{\yr}{\,\hbox{yr}}
\newcommand{\Myr}{\,\hbox{Myr}}
\newcommand{\Gyr}{\,\hbox{Gyr}}

\newcommand{\mJy}{\,\hbox{mJy}}
\newcommand{\muJy}{\,\hbox{\textmu{}Jy}}
\newcommand{\erg}{\,\hbox{erg}}
\newcommand{\K}{\,\hbox{K}}

\newcommand{\be}{\begin{equation}}
\newcommand{\ee}{\end{equation}}
\newcommand{\bea}{\begin{eqnarray}}
\newcommand{\eea}{\end{eqnarray}}

\def\la{~\raise.4ex\hbox{$<$}\llap{\lower.6ex\hbox{$\sim$}}~}
\def\ga{~\raise.4ex\hbox{$>$}\llap{\lower.6ex\hbox{$\sim$}}~}

\shorttitle{Cold debris disks}
\shortauthors{Krivov et al.}

\begin{document}



\title{
Herschel's ``Cold Debris Disks'':\\
Background Galaxies or Quiescent Rims of Planetary Systems?}


\author{{A. V. Krivov}}
\affil{Astrophysikalisches Institut und Universit{\"a}tssternwarte,
Friedrich-Schiller-Universit{\"a}t Jena,\\
Schillerg{\"a}{\ss}chen 2--3, 07745 Jena, Germany}

\email{krivov@astro.uni-jena.de}

\author{C. Eiroa}
\affil{Departamento de F\'isica Te\'orica, Facultad de Ciencias,
Universidad Aut\'onoma de Madrid,\\
Cantoblanco, 28049 Madrid, Spain}

\author{T. L\"ohne}
\affil{Astrophysikalisches Institut und Universit{\"a}tssternwarte,
Friedrich-Schiller-Universit{\"a}t Jena,\\
Schillerg{\"a}{\ss}chen 2--3, 07745 Jena, Germany}

\author{J. P. Marshall}
\affil{Departamento de F\'isica Te\'orica, Facultad de Ciencias,
Universidad Aut\'onoma de Madrid,\\
Cantoblanco, 28049 Madrid, Spain}

\author{B. Montesinos}
\affil{Departmento de Astrof\'isica, Centro de Astrobiolog\'ia (CAB, CSIC-INTA),\\
ESAC Campus, PO Box 78, 28691 Villanueva de la Ca\~nada, Madrid, Spain}

\author{C. del Burgo}
\affil{Instituto Nacional de Astrof\'isica, Optica y Electr\'onica (INAOE),\\
Aptdo. Postal 51 y 216, 72000 Puebla, Pue., Mexico}

\author{O. Absil}
\affil{Institut d'Astrophysique et de G\'eophysique, Universit\'e de
Li\`ege, all\'ee du 6 ao{\^u}t 17,
4000 Li\`ege, Belgium}

\author{D. Ardila}
\affil{NASA Herschel Science Center, California Institute of Technology,\\
1200 E. California Blvd., Pasadena, CA 91125, USA}

\author{J.-C. Augereau}
\affil{UJF-Grenoble 1 / CNRS-INSU, Institut de Plan\'etologie et
d'Astrophysique de Grenoble (IPAG),\\
UMR 5274, 38041 Grenoble, France}

\author{A. Bayo\altaffilmark{1}}
\affil{European Southern Observatory, Alonso de C\'ordova 3107, Vitacura, Santiago, Chile}
\altaffiltext{1}{Also at 
Max Planck Institut f\"ur Astronomie, K\"onigstuhl 17, 69117, Heidelberg, Germany}

\author{G. Bryden}
\affil{Jet Propulsion Laboratory, California Institute of Technology,
Pasadena, CA 91109, USA}

\author{W. Danchi}
\affil{NASA Goddard Space Flight Center, Exoplanets and Stellar Astrophysics,\\
Code 667, Greenbelt, MD 20771, USA}

\author{S. Ertel, J. Lebreton}
\affil{UJF-Grenoble 1 / CNRS-INSU, Institut de Plan\'etologie et
d'Astrophysique de Grenoble (IPAG),\\
UMR 5274, 38041 Grenoble, France}

\author{R. Liseau}
\affil{Department of Earth and Space Sciences, Chalmers University of Technology,\\
Onsala Space Observatory, 439 92, Onsala, Sweden}

\author{A. Mora}
\affil{ESA-ESAC Gaia SOC. PO Box 78 28691 Villanueva de la Ca\~nada,
Madrid, Spain}

\author{A. J. Mustill}
\affil{Departamento de F\'isica Te\'orica, Facultad de Ciencias,
Universidad Aut\'onoma de Madrid,\\
Cantoblanco, 28049 Madrid, Spain}

\author{H. Mutschke, R. Neuh\"auser}
\affil{Astrophysikalisches Institut und Universit{\"a}tssternwarte,
Friedrich-Schiller-Universit{\"a}t Jena,\\
Schillerg{\"a}{\ss}chen 2--3, 07745 Jena, Germany}

\author{G. L. Pilbratt}
\affil{ESA Astrophysics \& Fundamental Physics Missions Division,
ESTEC/SRE-SA,\\
Keplerlaan 1, 2201 AZ Noordwijk, The Netherlands}

\author{A. Roberge}
\affil{NASA Goddard Space Flight Center, Exoplanets and Stellar Astrophysics,
Code 667, Greenbelt, MD 20771, USA}

\author{T. O. B. Schmidt}
\affil{Astrophysikalisches Institut und Universit{\"a}tssternwarte,
Friedrich-Schiller-Universit{\"a}t Jena,\\
Schillerg{\"a}{\ss}chen 2--3, 07745 Jena, Germany}

\author{K. R. Stapelfeldt}
\affil{NASA Goddard Space Flight Center, Exoplanets and Stellar Astrophysics,\\
Code 667, Greenbelt, MD 20771, USA}

\author{Ph. Th\'ebault}
\affil{LESIA, Observatoire de Paris,
92195 Meudon Principal Cedex, France}

\author{Ch. Vitense}
\affil{Astrophysikalisches Institut und Universit{\"a}tssternwarte,
Friedrich-Schiller-Universit{\"a}t Jena,\\
Schillerg{\"a}{\ss}chen 2--3, 07745 Jena, Germany}

\author{G. J. White}
\affil{Department of Physics and Astrophysics, Open University, Walton
Hall, Milton Keynes MK7 6AA, UK}
\affil{Rutherford Appleton Laboratory, Chilton OX11 0QX, UK}

\and

\author{S. Wolf}
\affil{Christian-Albrechts-Universit\"at zu Kiel, Institut f\"ur
Theoretische Physik und Astrophysik,\\
Leibnizstr.  15, 24098 Kiel, Germany}

%
%
%



    
\begin{abstract}
Infrared excesses associated with debris disk host stars detected so
far, peak at wavelengths around $\sim 100\mum$ or shorter.
However, six out of 31 excess sources studied in the
{\it Herschel}\footnote{{\it Herschel}
is an ESA space observatory with science instruments
provided by European-led Principal Investigator consortia and with important
participation from NASA.}\ 
Open Time Key
Programme, DUNES, have been seen to show
significant~-- and in some cases extended~-- excess emission at $160\mum$,
which is larger than the $100\mum$ excess.
This excess emission has been attributed to circumstellar dust
and has been suggested to stem from 
debris disks colder than those known previously.
Since the excess emission of the cold disk
candidates is extremely weak, challenging even the unrivaled sensitivity
of {\it Herschel}, it is prudent to carefully consider whether some or even
all of them may represent unrelated galactic or
extragalactic emission, or even instrumental noise.
We re-address these issues using several distinct methods and conclude that 
it is highly unlikely that none of the candidates  
represents a true circumstellar disk.
For true disks,
both the dust temperatures inferred from the spectral energy distributions
and the disk radii estimated from the images suggest that the dust is
nearly as cold as a blackbody.  This requires
the grains to be larger than $\sim 100\mum$,
even if they are rich in ices or are composed of any other
material with a low absorption in the visible. The dearth of
small grains is puzzling, since collisional models of debris disks
predict that grains of all sizes down to several times the radiation
pressure blowout limit should be present. We explore several conceivable
scenarios: transport-dominated disks, disks of low dynamical excitation,
and disks of unstirred primordial macroscopic grains.
Our qualitative analysis and collisional simulations rule out the
first two of these scenarios, but show the feasibility of the third one.
We show that such disks can indeed survive for gigayears, largely preserving
the primordial size distribution. They should be composed of macroscopic
solids larger than millimeters, but smaller than a few kilometers in size.
If larger planetesimals were present, they would stir the disk, triggering
a collisional cascade and thus causing production of small debris, which is not seen.
Thus planetesimal formation, at least in the outer regions of the systems, has 
stopped before ``cometary'' or ``asteroidal'' sizes were reached.
\end{abstract}



\section{Introduction}

A significant fraction of main-sequence stars are found to be surrounded by
detectable belts of debris, composed of planetesimals and their dust
\citep[see, e.g.][for recent reviews]{wyatt-2008,krivov-2010}.
This material, together with planets, must represent the natural remnants
of the systems' evolution during the protoplanetary phase.
The observed
debris disks typically reside on the outskirts of their host systems,
just as for the solar system's Kuiper belt. The vast majority of
these disks have been detected by observation of the thermal emission
from their constituent dust at far-infrared (IR) wavelengths at levels
above those predicted for the stellar photospheres. Several tens
of these systems have been resolved (in at least one axis) providing
crucial measurement of the dust spatial location around their host
stars, though the bulk remain unresolved and can only be interpreted
through their point-like thermal emission.

One of the serendipitous results of the {\it Herschel} Open Time Key Programme
(OTKP) DUNES \citep{eiroa-et-al-2010}
has been a tentative identification of a new class of ``cold'' debris disks
\citep{eiroa-et-al-2011}.
These are the cases where sources
show a significant IR excess at $160\mum$ and possibly also at longer wavelengths,
but a smaller or no IR excess at $100\mum$.
This contrasts with all of the debris disks observed previously,
whose thermal emission peaks at wavelengths $\la 100\mum$.
Since the excess emission of the cold disk candidates is extremely weak,
significant effort 
has gone into assessing the possible contamination
by galactic or extra-galactic emission, or even instrumental 
noise and to show that the excess emission is likely
to originate from the circumstellar dust.

Another {\it Herschel} OTKP, H-ATLAS, has reported two debris disk candidates with
similar properties \citep{thompson-et-al-2010}.
Their identification was based on the excess detection
at $250\mum$ and $350\mum$ and upper limits at shorter
and longer wavelengths.
Rough estimates show that these may have similar properties as the DUNES
cold disk candidates, except for substantially higher dust masses.
Unfortunately, both candidates are located at large distances
of $\sim 100$-$200\pc$. As a result, the upper limits on the excess
fluxes at $100\mum$ and $160\mum$ are very weak, as are the upper limits
on the disk radii. This renders the formation of any definite conclusions
extremely difficult,
so that the H-ATLAS candidates are not further discussed here.

On the assumption that the DUNES ``cold disks'' are true disks,
the inferred dust temperatures are as low as $T_d \sim 20$--$30\K$
\citep{eiroa-et-al-2011}.
Even though these disks have low fractional luminosities
($f_d \sim 10^{-6}$), some of them reveal extended emission, allowing one
to roughly estimate how far from the star the dust is located.
The dust temperatures derived from the spectral energy distributions (SEDs)
turn out to be comparable to the blackbody temperatures at the dust location.
For those sources with point-like emission,
an upper limit on the disk radius can be placed,
constraining the maximum distance from the star,
where dust is still allowed to be for the disk to remain unresolved.
In these cases, too, the observed dust emission is nearly as cold
as the blackbody emission at that distance.

Explaining how the material in true cold disks can be as cold as inferred
is a challenge.
The nearly blackbody temperatures of grains are suggestive of large grains
(compared to far-IR wavelengths).
It is not clear whether the requirement of large sizes can be relaxed
by assuming dust compositions with a low absorption in the visible,
such as icy grains.
The lack of small grains would be difficult to explain,
because interpretation of multiwavelength resolved images
of numerous debris disks done so far reveals dominating sizes to lie
in the micrometer range
\citep[see, e.g.,][and references therein]{wyatt-2008,krivov-2010}.
This is also in accord with collisional models of debris disks that
robustly predict all of the grains
down to several times the radiation pressure blowout limit to be present
\citep[e.g.][]{wyatt-et-al-1999, wyatt-dent-2002, krivov-et-al-2006,thebault-augereau-2007}.
Note that all of the DUNES targets are solar-type (FGK) stars,
for which the blowout radius~-- the radius of the smallest grains that can stay
in bound orbits around the star against the radiation pressure~-- is $\la 1\mum$.

This paper extends the previous analysis
\citep{eiroa-et-al-2011} from three
to all six cold disk candidates identified by DUNES;
revisits the hypotheses of unrelated sources and ``false alarms'';
and assuming that one or more of the candidates are real disks,
explores possible reasons for the cold disk phenomenon.
In Section~2, we describe a selection of cold disk candidates
and re-address the disambiguation issue with possible background galaxies.
In Section~3, we constrain the properties of the emitting material,
trying to find grain sizes and compositions that are compatible with the
observational data.
In Section~4, we attempt to find an appropriate physical regime for
debris disks that would provide the grains with the properties found
in Section~3.
Section~5 contains our conclusions and a discussion and
Section~6 offers a short summary of our findings.

\section{Cold Disk Candidates}

\subsection{Characterization of candidates}

The OTKP DUNES \citep{eiroa-et-al-2010}\footnote{see also 
http://www.mpia-hd.mpg.de/DUNES/.} is a survey of nearby
($d \la 20$--$25\pc$) Sun-like (FGK) stars that used
{\it Herschel} \citep{pilbratt-et-al-2010}
PACS \citep{poglitsch-et-al-2010} and
SPIRE \citep{griffin-et-al-2010} scan map observations in six
broad photometric bands
around $70$, $100$, $160$, $250$, $350$, and $500\mum$.
The DUNES strategy has been to integrate as deeply as needed
to reach the photospheric level (at $100\mum$) of all of the sources,
facilitating detection of faint excess emission.
In the entire DUNES dataset of 133 stars we have identified 31
sources as having a significant excess at least at one of these wavelengths
\citep{eiroa-et-al-2013}.
Six out of these 31 have been classified as cold disk candidates,
showing a significant (at least $>3\sigma$) excess at $160\mum$ that exceeds
the excess at $100\mum$ (if the latter is present at all).
Three objects (HIP~29271, HIP~49908, and HIP 109378)
have been discussed in more detail in \citet{eiroa-et-al-2011}.
The three additional candidates are HIP~171, HIP~73100, and HIP~92043.

\begin{table*}[htb!]
\begin{center}

  \caption{Stellar Parameters \label{tab:six_parms}}

\begin{tabular}{rrclrrrcclrcc}
\tableline\tableline
HIP	& HD	& Name		& Sp	& $d$, pc	& b	&$\beta$& $L/L_\odot$	& $T_\mathrm{eff}$, K	& log g	& [Fe/H]& Age, Gyr	& Ref.		\\
\tableline
   171A	&224930	& 85 Peg A	& G5V	& 12.17		&$-34.5$&$+24.5$& 0.613		& $5600$		& 4.51	& -0.66	& 3.1, 4.0	& 1		\\
      B	& 	& 85 Peg B	& K7V	& 12.17		&	&	& 0.074		& $4200$		& 4.77	& -0.64	&		& 1		\\
 29271	& 43834	&$\alpha$~Men	& G5V	& 10.20		&$-28.8$&$-81.8$& 0.847		& $5591$		& 4.46	&  0.08	& 3.4, 5.5	& 2		\\ 
 49908	& 88230	&		& K8V	&  4.87		&$+52.1$&$+35.4$& 0.125		& $4081$		& 4.71	& -0.16	& 3.2, --	& 3,4		\\ 
 73100	&132254	&		& F7V	& 25.11		&$+57.2$&$+61.4$& 2.831		& $6220$		& 4.15	& -0.03	& 1.2, 7.2	& 2		\\ 
 92043	&173667	&110~Her	& F6V	& 19.21		&$+10.4$&$+43.4$& 6.141		& $6431$		& 4.08	&  0.04	& 0.3, 4.7	& 2		\\ 
109378	&210277	&		& G0V	& 21.56		&$-46.9$&$ +3.6$& 1.002		& $5540$		& 4.39	&  0.22	& --, --	& 2		\\ 
\tableline
\end{tabular}

\end{center}

  \tablecomments{
  $b$ and $\beta$ are the galactic and the ecliptic latitude in degrees, respectively.
  The age column lists two values: from X-ray luminosity and based on the R$^\prime$(HK)-index.}

  \tablerefs{
  (1) \citet{bach-et-al-2009};
  (2) \citet{eiroa-et-al-2013};
  (3) \citet{boyajian-et-al-2012};
  (4) \citet{anderson-francis-2011}.
  }

\end{table*}

The observations, data reduction procedures, photometry calculations,
and the method of computing
the expected photospheric fluxes at PACS and SPIRE wavelengths
for all sources in the DUNES sample are described in detail
in \citet{eiroa-et-al-2013}.
ObsID for the PACS and SPIRE observations are also listed there.
Nonetheless, for the reader's convenience, here we
give some details of the {\it Herschel} observations, data reduction,
photometric estimates, and photospheric predictions specific to the six stars in question.

PACS scan  map observations consisted of 10 legs of
3$\arcmin$ length, with a 4$\arcsec$ separation between legs, scanning
at the  medium slew speed  (20$\arcsec$ per second).  Each  target was
observed   at   two  array   orientation   angles  (70$^{\circ}$   and
110$^{\circ}$)  to  improve noise  suppression  and  to  assist in  the
removal  of low  frequency ($1/f$)  noise, instrumental  artifacts and
glitches from  the images. The SPIRE observation of HIP 92043~consisted of
five repeats of the small scan map mode\footnote{see
http://herschel.esac.esa.int/Docs/SPIRE/html/spire\_om.pdf for details.},
producing  a  fully  sampled  map   covering  a  region
4$\arcmin$ around  the target. PACS  and SPIRE  observations  were  reduced
using the  Herschel Interactive Processing Environment, HIPE \citep{ott-2010},
user release version 7.2, PACS calibration version 32 and SPIRE calibration
version 8.1. The individual PACS scans  were processed with a high pass filter
to remove  background structure,  using high pass  filter radii  of 15
frames  at  $70\mum$,  20  frames  at $100\mum$,  and  25  frames  at
$160\mum$, suppressing structure larger than 62$\arcsec$, 82$\arcsec$
and 102$\arcsec$ in the  final images, respectively. For the filtering
process,  regions of  the map  where the  pixel brightness  exceeded a
threshold defined as twice the standard deviation of the non-zero flux
elements in the map were masked from inclusion in the high pass filter
calculation.   Deglitching  was carried  out  using  the second  level
spatial deglitching  task, following issues  with the clipping  of the
cores of  bright sources  using the MMT  deglitching method.   The two
individual PACS scans were mosaicked  to reduce sky noise and suppress
$1/f$  stripping effects from  the scanning.   Final image  scales were
1$\arcsec$ per pixel at $70\mum$  and $100\mum$ and 2$\arcsec$ per pixel at
$160\mum$ compared to native instrument pixel sizes of 3$\farcs$2 and
6$\farcs$4. For  the SPIRE observation,  the small map  was created
using the  standard pipeline routine  in HIPE, using the  naive mapper
option.  Image  scales of 6$\arcsec$, 10$\arcsec$  and 14$\arcsec$ per
pixel were used at $250\mum$,  $350\mum$, and  $500\mum$.

PACS photometry  was  carried out using two different methods. The
first   method consisted   in   estimating   PACS  fluxes
using  circular aperture  photometry with radii
4$\arcsec$, 5$\arcsec$,  and 8$\arcsec$ at $70\mum$, $100\mum$, and  
$160\mum$,  respectively. These apertures were chosen in the case of point
sources and because they provide the highest SNR as estimated by the {\it Herschel}
team\footnote{Technical Note PICC-ME-TN-037 in
http://herschel.esac.esa.int.} and confirmed by our own analysis \citep{eiroa-et-al-2013}.
The corresponding  beam  aperture correction  as  given  in the
{\it Herschel} technical note PICC-ME-TN-037 was taken into account.
The reference background region was usually taken as a ring  of width
10$\arcsec$ at a separation of 10$\arcsec$  from the  circular  aperture size.
Nonetheless we took
special  care to  choose the  reference sky  region for  those objects
where the ``default''  sky was or could be  contaminated by background
objects, e.g., in the cases of HIP~171 and HIP~109378. Sky noise for each PACS
band was calculated from the rms pixel variance of ten sky apertures of
the same size as the source aperture
and randomly distributed  across the  uniformly covered  part  of the
image.  Final  error estimates  take into account  a correlated
noise factor of 3.7, as estimated by us for the DUNES observations, which
is a bit larger than the one given in the technical note PICC-ME-TN-037
\citep[see][]{eiroa-et-al-2013}. The second method used to estimate the photometry
consisted in using rectangular  boxes  with  areas  equivalent to  the
default  circular apertures. However, for extended sources (HIP~29271,
HIP~49908 and HIP~92043 at $160\mum$),  we chose boxes large enough to
cover the region where the emission
is significant as compared to the background noise.
The sky level and sky rms noise for
this method  were estimated from measurements in ten fields, selected
as clean as possible by the eye, of the same  size as the photometric
source boxes.   Photometric values and  errors take into  account beam
correction factors.  The estimated  fluxes from both methods, circular
and  rectangular aperture  photometry, agree  within the  errors.
The SPIRE $250\mum$ flux of HIP~92043 was estimated using the SUSSEXtractor tool.

The synthetic stellar spectra of the six stars
were calculated with the PHOENIX/GAIA models \citep{brott-hauschildt-2005},
using the stellar parameters given
in Tab.~\ref{tab:six_parms}.
These spectra were fitted vertically to the optical and near-IR photometry
obtained from all available sources.
These included
{\it Hipparcos} catalog (B, V, I),
the Str\"omgren u, v, b, y photometry \citep{hauck-mermilliod-1998},
UKIRT-B (V, J, K, L$^\prime$),
2MASS \citep[J, H, K$_s$; see][]{cutri-et-al-2003,skrutskie-et-al-2006} with quality flags~A and B,
JP11 catalog (J, H, K, L),
and {\it WISE} W1, W3, and W4 bands (W2 was excluded because of 
significant flux calibration problems).
One of the stars, HIP~171 (=85 Peg), needed a special treatment.
It is a spectroscopic binary
(distance 12.17 pc, separation between both components $0.83\arcsec$ or $10.1\AU$,
period 26.3 yr, and eccentricity 0.38), and
a photospheric fit considering both components A and B was done\footnote{In fact,
HIP~171~B is likely to have an additional companion Bb
with a mass $\sim 0.1$--$0.2 M_\odot$
\citep[see][and references therein]{bach-et-al-2009}.
Furthermore, \citet{schmitt-1997} has suggested the presence of an additional component C
more than $1^\prime$ away with spectral type M6 and $V=17$
(see also Washington Double Star catalog).}.
For that star, {\it WISE} W4 was not used because there are indications that another,
warm excess that starts around W4 may be present \citep{koerner-et-al-2010}. 

The method of normalization of the photospheric models to the photometry is explained in 
detail in App.~C of \citet{eiroa-et-al-2013}.
Five subsets of the full SED were chosen to carry out five normalizations, namely, VI+nIR, 
BVI+nIR, VI+nIR+WISE, nIR+WISE and VI+nIR+WISE \citep[see][]{eiroa-et-al-2013}.
For each of them, a reduced $\chi^2$ was computed.
The normalization with the least reduced $\chi^2$ was then selected and used to
predict the photospheric fluxes S$_\lambda$ at the PACS and SPIRE wavelengths.

\begin{table*}[htb!]
\begin{center}

  \caption{Fluxes and Disk Parameter Estimates   \label{tab:six_data}}

\begin{tabular}{lcccccc}
\tableline\tableline
   HIP                                   & 171\tablenotemark{a,c}    & 29271\tablenotemark{b}    & 49908\tablenotemark{b,d} & 73100\tablenotemark{a,e}   & 92043\tablenotemark{b,f}   & 109378\tablenotemark{a}   \\
\tableline
   S$_{70}$                              & ---                       & ---                       & ---                      & $14.4\pm0.2$               & $48.8\pm0.8$               & ---                       \\
   F$_{70}$                              & ---                       & ---                       & ---                      & $24.7\pm 3.2$              & $59.0\pm3.5$               & ---                       \\
   $\chi_{70}$                           & ---                       & ---                       & ---                      & $10.3\pm 3.2$              & $10.2\pm3.6$               & ---                       \\
   $\chi_{70}/\sigma_{70}$               & ---                       & ---                       & ---                      & 3.2                        & 2.8                        & ---                       \\
\tableline
   S$_{100}$                             & $11.4\pm0.2$              & $17.4\pm0.2$              & $24.6\pm1.0$             & $ 7.1\pm0.1$               & $23.9\pm0.4$               & $4.7\pm0.1$               \\
   F$_{100}$ (PACS)                      & $11.7\pm1.3$              & $17.8\pm1.3$              & $22.5\pm0.9$             & $13.8\pm0.8$               & $30.2\pm2.4$               & $8.5\pm1.0$               \\
   $\chi_{100}$                          & ---                       & ---                       & ---                      & $ 6.7\pm0.8$               & $ 6.3\pm2.4$               & $3.8\pm1.0$               \\
   $\chi_{100}/\sigma_{100}$             & ---                       & ---                       & ---                      & 8.4                        & 2.6                        & 3.8                       \\
\tableline
   S$_{160}$                             & $ 4.4\pm0.09$             & $ 6.8\pm0.08$             & $ 9.6\pm0.40$            & $ 2.8\pm0.04$              & $ 9.3\pm0.16$              & $ 1.8\pm0.03$             \\
   F$_{160}$ (PACS)                      & $12.5\pm2.4$              & $14.4\pm2.0$              & $16.0\pm1.7$             & $12.5\pm1.7$               & $21.9\pm3.8$               & $12.4\pm1.6$              \\
   $\chi_{160}$                          & $ 8.1\pm2.4$              & $ 7.6\pm2.0$              & $ 6.4\pm1.7$             & $ 9.7\pm1.7$               & $12.6\pm3.8$               & $10.6\pm1.6$              \\
   $\chi_{160}/\sigma_{160}$             & 3.4                       & 3.8                       & 3.8                      & 5.7                        & 3.3                        & 6.6                       \\
\tableline
   S$_{250}$                             & ---                       & ---                       & ---                      & ---                        & $ 3.8\pm0.07$              & ---                       \\
   F$_{250}$ (SPIRE)                     & ---                       & ---                       & ---                      & ---                        & $11.1\pm7.2$               & ---                       \\
   $\chi_{250}$                          & ---                       & ---                       & ---                      & ---                        & $ 7.3\pm7.2$               & ---                       \\
   $\chi_{250}/\sigma_{250}$             & ---                       & ---                       & ---                      & ---                        & 1.0                        & ---                       \\
\tableline
   $R_\mathrm{disk}$                     & $< 6\arcsec$ $ (<70\AU)$  & $  8\arcsec$ $  (80\AU)$  & $ 10\arcsec$ $  (50\AU)$ & $< 6\arcsec$ $(<150\AU)$   & $  7\arcsec$ $( 130\AU)$   & $< 6\arcsec$ $(<130\AU)$  \\
   $f_d$                                 & $3 \times 10^{-5}$        & $1 \times 10^{-6}$        & $2 \times 10^{-6}$       & $3 \times 10^{-6}$         & $5 \times 10^{-7}$         & $5 \times 10^{-6}$        \\
\tableline
   Ref                                   & 1                         & 1, 2                      & 1, 2                     & 1                          & 1                          & 1, 2                      \\
\tableline
\end{tabular}

\end{center}

  \tablecomments{
  Predicted photospheric (``S''), observed (``F''), and excess (``$\chi$'') fluxes
  and their 1-$\sigma$ uncertainties  (in mJy),
  significance of the excesses ($\chi/\sigma$), 
  disk radius estimates or upper limits
  inferred from the analysis of the images ($R_\mathrm{disk}$),
  and the dust fractional luminosity ($f_d$).}

  \tablerefs{
  (1) \citet{eiroa-et-al-2013};
  (2) \citet{eiroa-et-al-2011}.}

  \smallskip

  \tablenotetext{1}{Point-like both at $100\mum$ and $160\mum$}
  \tablenotetext{2}{$R_\mathrm{disk}$ is an estimate from the deconvolved $160\mum$ image}
  \tablenotetext{3}{The only excess source with mispointing around $2\sigma$ ($5.5\arcsec$).
      An additional check of the accuracy of the predicted photospheric fluxes
      was done by using the radii of components A and B
      given in \citet{bach-et-al-2009},
      $0.834$ and $0.512 R_\odot$, and the distance, $d = 12.17\pm 0.33\pc$,
      from the {\it Hipparcos} parallax $82.17 \pm 2.23$~mas \citep{vanleeuwen-2007}.
      This enabled us to compute the `dilution factors' $(R_\star/d)^2$ for each 
      star.
      The individual PHOENIX model photospheres for A and B were multiplied by the
      corresponding dilution factors to obtain the flux density measured at Earth,
      and then added to build the composite final photospheric SED.
      The uncertainty in each dilution factor was estimated by propagating the
      errors according to the individual uncertainties in $R_\star$ (not given
      in Bach et al., we assumed it to be $5\%$) and $d$ (as indicated above).
      With this, the predictions for S$_{100}$ and S$_{160}$ are
      $11.6 \pm 1.2\mJy$  and $4.5 \pm 0.5\mJy$, respectively.
      This is in an excellent agreement with the values given in the Table
      }
  \tablenotetext{4}{In a similar manner as the one described in note (c)
      the very accurate 
      stellar radius ($0.6415\pm 0.0048 R_\odot$) given by \citet{boyajian-et-al-2012}
      for HIP 49908 and the distance $4.87\pm 0.01\pc$ from the {\it Hipparcos}
      parallax $205.21\pm 0.54$\,mas \citep{vanleeuwen-2007} were used to compute the
      dilution factor $(R_\star/d)^2$.
      The PHOENIX model photosphere was scaled by
      the dilution factor to build the photospheric SED.
      The uncertainties in the radius and distance were used to estimate the uncertainty
      in the dilution factor. The results for the predicted fluxes are
      S$_{100} = 24.8 \pm 0.4\mJy$ and S$_{160} = 9.7 \pm 0.15\mJy$,
      i.e.~within $0.2\mJy$ from the values listed in the Table
      }
  \tablenotetext{5}{F$_{70}$ from MIPS (reduction by G. Bryden)}
  \tablenotetext{6}{F$_{70}$ from PACS.
      The MIPS measurement gives $69.8\pm 8.9\mJy$ (reduction by G. Bryden)}

\end{table*}



The predicted photospheric fluxes together with the
{\it Herschel} measurements are listed in Tab.~\ref{tab:six_data}.
The indicated photospheric uncertainties should be taken as lower limits,
because these do not include the individual uncertainties in each observed
point.
However, to realistically estimate the actual accuracy of the photospheric
predictions, we made additional checks.
For two stars, HIP 171 and HIP 49908,
we carried out a scaling of the photospheric models with a 
completely different method, independent of the optical and near-IR photometry.
We took advantage of the fact that for these objects the stellar 
radii, $R_\star$, had been measured with high accuracy.
Since the distance to the stars, $d$, is also known,
this enabled us to compute the `dilution factors' $(R_\star/d)^2$ for each star.
The PHOENIX/GAIA model photospheres provide
the flux density at the stellar surface.
Multiplying it by the dilution factor,
one obtains directly the flux density measured at Earth.
The predictions of the photospheric fluxes obtained in this way match very 
well those from the method described above
(see notes c and d to Tab.~\ref{tab:six_data}).

Apart from the fluxes, Tab.~\ref{tab:six_data} gives an estimate of the disk radius
$R_\mathrm{disk}$. If the emission at $160\mum$ is extended, we derive it from the deconvolved
image of the source. If it appears point-like, we give an upper limit on $R_\mathrm{disk}$
of 1/2~FWHM at $160\mum$ (as indicated by the ``$<$'' sign).

\begin{figure*}[htb!]
  \begin{center}
  \includegraphics[width=0.77\textwidth]{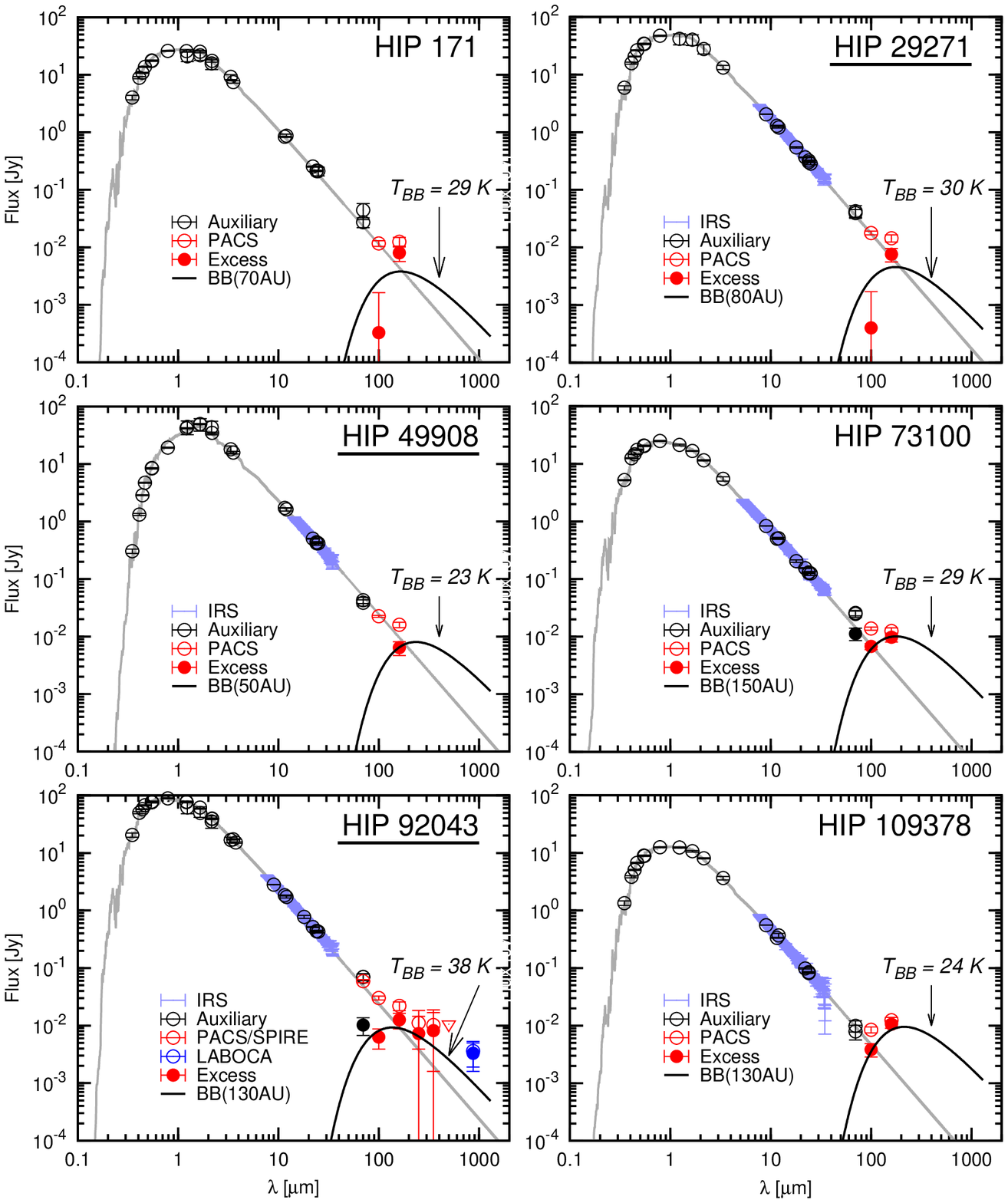}
  \end{center}
  \caption{
  SEDs of six cold disk candidates
  (names of the resolved ones are underlined).
  Each plot presents auxiliary photometry (black symbols)
  and, where available, the {\it Spitzer}/IRS spectrum
  (light blue line whose thickness reflects error bars).
  The preferred photosphere is depicted with grey line.
  Superimposed are PACS and~-- where available~--  SPIRE data (red)
  and LABOCA (blue) points.
  Open symbols are total fluxes, filled ones are excess fluxes
  (i.e., photosphere-subtracted).
  All symbols are with 1-$\sigma$ error bars, which are often smaller than the symbols.
  Circles are detections, triangles are 1-$\sigma$ upper limits.
  Overplotted with a black line is emission from blackbody dust,
  placed at the distances $R_\mathrm{disk}$ given in Table~\ref{tab:six_data}.
  }
  \label{fig:six}
\end{figure*}

Figure~\ref{fig:six} displays the SEDs.
It shows all ancillary photometry points, including
those not used for the photospheric normalization, such as
the {\it AKARI} $9\mum$ and $18\mum$ fluxes and
{\it Spitzer}/MIPS $24\mum$ and $70\mum$ data.
The best-fit photospheric models are overplotted.
The same figure depicts our {\it Herschel}/PACS and /SPIRE measurements
and upper limits.
In addition, it plots the excess fluxes, obtained after the photospheric subtraction.

\subsection{Origin of the observed emission}

Table~\ref{tab:six_data} shows that the significance of the
$160\mum$ excess detection is larger than $5\sigma$ for
two sources, HIP~73100 ($5.7\sigma$) and HIP~109378 ($6.6\sigma$),
while the other four sources only have a significance
between $3.3\sigma$ and $3.8\sigma$.
At this level of confidence, there remains a possibility that
some of the cases are false detections due to noise,
with a gaussian probability ranging from
$4\times 10^{-11}$ for HIP~109378 to $1 \times 10^{-3}$ for HIP~92043.

Note that the probability that these stars do not have {\em any}
far-IR excess at all is much lower, since some of them show
excess in the bands other than $160\mum$.
Let $P_i = 1 - \mathrm{erf}((\chi_i/\sigma_i)/\sqrt{2})$ be the probability that the 
detection in band $i$ with a significance $\chi_i$ is false.
For detections in multiple bands $i = 1, ..., n$ with
significances $\chi_i$, the probability that all of them are false is
\be
P_\mathrm{comb} = 1 - \mathrm{erf}\left(\chi_\mathrm{comb}/\sigma \over \sqrt{2} \right)
         = \prod_{i=1}^n P_i(\chi_i) ,
\label{eq:comb}
\ee
which is an equation to solve for the combined significance
in units of the standard deviation, $\chi_\mathrm{comb}/\sigma$.
For instance, HIP~92043  has  a 3.3-$\sigma$ excess at $160\mum$,
a 2.6-$\sigma$ excess at $100\mum$,
and a 2.8-$\sigma$ excess at $70\mum$.
From Eq.~(\ref{eq:comb}), the combined three-band excess significance level
is $5.5\sigma$,
and the gaussian probability that the entire emission of that star
is purely photospheric is $4\times 10^{-8}$.
%

While it is extremely unlikely that the excess emission is just noise,
the SEDs shown in Fig.~\ref{fig:six}, as well as the fact that the DUNES 
fields
are very deep (with a PACS~100/160 on-source time of up to $1440\s$)
raise the question of whether some of the
cold disk candidates, or even all of them, may be associated with
the galactic background radiation or extragalactic background
rather than represent the true circumstellar emission.
We deem the former possibility unlikely.
Although diffuse cirrus bands are clearly
seen around the position of some of the DUNES sources
\citep{eiroa-et-al-2013},
there is no contamination of this kind in the fields of our six candidates,
all of which lie more than $10^\circ$ above or below the galactic plane.
An unrelated Milky Way object in the line of sight, such as a cold free-floating dwarf,
would have to be closer than 1~pc to
produce a $30\K$ emission at the same flux level as the one observed.
A probability of having a transneptunian object in the foreground
even in one of the six cases is very small, and for all six just negligible.
Note that all of the sources, except for HIP~109378, are more than $24^\circ$
from the ecliptic
(see Tab.~\ref{tab:six_parms}).

In contrast, background galaxies pose a serious difficulty.
Extragalactic sources unrelated to the targets are clearly visible in most of
the fields observed by DUNES, and the SEDs of our six sources peaking 
longward of $160\mum$ would not be untypical of moderately red-shifted
galaxies. 
Even some of the bright, well resolved disks observed by {\it Herschel} were found to
be contaminated, as exemplified by 61~Vir \citep{wyatt-et-al-2012}.
In the DUNES sample of 133 stars, apparent excess emission of seven sources is likely to
derive entirely from the extragalactic background
\citep[][their App.~D]{eiroa-et-al-2013}.

Ruling out a possible confusion solely by the shape of SEDs is not possible.
Taking into account a spread of redshifts and luminosities of potential
extragalactic contaminators,
one can always attribute a few photometric points to a background galaxy,
especially given the uncertainties in the measured fluxes.
Alternatively, one could use the morphology of the extended emission to distinguish between
the true disks and background sources.
Pronounced asymmetry of the emission and a strong offset from the stellar position
might favor a background object.
Unfortunately, the resolution of {\it Herschel}
at far-IR wavelengths, low surface brightness of the cold disk candidates,
and complicated background patterns around their position do not allow any
definite conclusion on the shape of the observed emission (which appears extended
at $160\mum$ for HIP~29271, HIP~49908 and HIP~92043).
Even a perfectly symmetric disk centered on the star may easily appear highly asymmetric in 
observations with low signal-to-noise ratios (SNR).
This may be best illustrated by the fact that even a bright, pole-on Vega disk exhibited a 
bright blob on one side of the star when viewed early with SCUBA \citep{holland-et-al-1998}, 
which was neither confirmed later by millimeter observations with higher SNR and better 
resolution \citep{hughes-et-al-2012} nor by {\it Spitzer} \citep{su-et-al-2005}
and {\it Herschel} \citep{sibthorpe-et-al-2010} observations at shorter wavelengths.  

The most direct and reliable way of disentangling the possible confusion towards
the cold disk candidates would be to take a second epoch with PACS, trying to
find out whether the excess $160\mum$ emission has moved with respect to the $100\mum$ one
and the optical position of the star.
Unfortunately, all of these observations
presented here are single-epoch ones and, even if taking the second epoch were 
possible,
the lifetime of the {\it Herschel} mission of less than 4 years would not be long enough
to exclude the background hypothesis, given the proper motion of our candidates
(the largest, that of HIP~49908, is $1.4\arcsec/\yr$).

We therefore tried yet another method, which was to search the fields
around the optical positions of the six stars in catalogs and archives
for possible extragalactic objects.
For X-ray sources, we accessed the
{\it XMM}, {\it Chandra}, and {\it ROSAT} data,
using NASA's HEASARC system\footnote{http://heasarc.gsfc.nasa.gov/.}.
The diameter of the search fields was taken to be $\approx 2^\prime$,
because of the rather low positional accuracy of the
X-ray instruments (e.g., up to $96\arcsec$ for faint {\it ROSAT} sources),
so that any object in such a field
could be an X-ray counterpart to the {\it Herschel} source.
Detection of X-ray sources that are inconsistent in their flux with the stars themselves
could indicate background galaxies that potentially also contaminate our measurements in the
far-IR.
Where possible, we have also searched the same fields for
optical and near-IR
counterparts, using deep images from the HST and VLT/NaCo archives
and digitized plate scans from the USNO-A and DSS
(Digitized Sky Survey) archives.

{\em HIP~171}.
The {\it ROSAT} catalog \citep{voges-et-al-2000}
contains an X-ray source (1RXS~J000211.4+270515)
with a nominal position $30\arcsec$ north of the star and a count
rate of $0.029 \pm 0.011$~ct/s.
The All-Sky Optical Catalog of Radio/X-Ray
Sources \citep{flesch-hardcastle-2004} reports an optical counterpart
($R=20$, $B=22$) to that source with a 69\% likelihood of being a galaxy. 
However, the digitized plates on which this identification is based are
heavily contaminated by the bright nearby star (V=5.75 mag),
hampering a detection of such a faint galaxy.
Visual inspection suggests that the star HIP 171~-- being very bright,
overexposed and saturated~-- was misclassified as an extended source and,
hence, misidentified as a galaxy.
Our interpretation is fully consistent with the study of \citet{schmitt-1997},
who finds that the X-ray emission of the X-ray source
1RXS J000211.4+270515 is typical for a late-type star
and concludes that either the spectroscopic binary HIP~171~A$+$B itself
or the wide separated later-type companion emits the X-rays.
He specifically points out that there
is no evidence for an additional background source.

\begin{figure}[htb!]
  \begin{center}
  \includegraphics[width=0.48\textwidth]{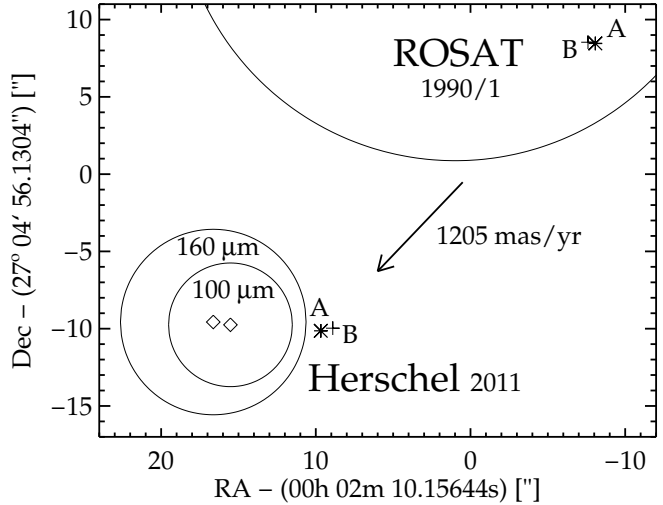}
  \end{center}
  \caption{
Astrometry of HIP~171. All positions are given in ICRS 2000 relative
to the catalog position of the star in 2000.
Right upper corner: calculated positions of the components A (asterisk) and B (cross)
at the moment of {\it ROSAT} observations
(between 1990-07-15 and 1991-01-16).
Left bottom corner: their positions at the moment of the {\it Herschel} observation
on 2011-01-17.
Since the exact epoch of the {\it ROSAT} observation is not known,
the proper motion over 6 months can introduce
a positional uncertainty of $0.6\arcsec$.
The uncertainty in the position due to that of the proper motion itself
(52 mas in 20 years) and the parallactic wobble (82 mas)
are smaller than the size of the asterisk.
{\it ROSAT} has detected an X-ray source at
$00^h02^m11.40^s$ and $+27^\circ05'15.0"$
($+1.0\arcsec$,$+18.9\arcsec$ in the relative coordinates 
used here; lies outside the plot).
A circle with a radius of $18\arcsec$ around that point
represents a 1-$\sigma$ positional accuracy of {\it ROSAT} in this case.
The calculated position of the star at the moment of the {\it ROSAT} observation
coincides with the X-ray source position within $0.77\sigma$.
The diamonds in the left bottom part of the plot show the centers
of the $100\mum$ and $160\mum$ emission measured by {\it Herschel} in 2011,
the circles around them correspond to the FWHMs at those wavelengths.
The calculated positions of the star in 2011 and the emission measured by
{\it Herschel} are consistent with each other to nearly 1-$\sigma$.
  }
  \label{fig:HIP171}
\end{figure}

It should also be noted that the {\it ROSAT} observation dates back to
1990, whereas the {\it Herschel} data were taken in 2011.
Given the proper motion of the star
($780 \pm 2$\,mas/yr in RA $\cos$(Dec)
and $-918 \pm 1$\,mas/yr in Dec), it has moved
between the two observations by $24\arcsec$ towards southeast (PA$=140^\circ$),
explaining much of the $30\arcsec$ offset mentioned above
(Fig~\ref{fig:HIP171}).

{\em HIP~29271} ($\alpha$~Men).
The star, which also has a close M-dwarf true companion at $30\AU$ 
\citep{eggenberger-et-al-2007}, has been
associated with an X-ray source (2RXP~J061014.3-744510, {\it ROSAT} count rate:
$0.024\pm0.003$~ct/s).

{\em HIP~49908}.
HIP~49908 is a flare star that has been detected as an X-ray source, too.

{\em HIP~73100}.
The star is also seen in X-rays 
(1RXS~J145619.7+493753, $0.027\pm0.009$~ct/s).
The DSS image around HIP~73100 reveals some galaxies that match bright neighboring 
sources in the PACS field.
However, the HST archive image is clean within $8\arcsec$ from the star.

{\em HIP~92043}.
HIP~92043 has been detected as a strong X-ray source
(2RXP~J101122.8+492714, $0.16$~ct/s).

{\em HIP~109378}.
No X-ray source associated with HIP~109378 has been found.
In the near-IR, \citet{eggenberger-et-al-2007} report no additional sources
around HIP~109378 from their VLT/NaCo narrow-band data.

In summary, there is an X-ray source in five
out of six fields, which is most likely associated
with the star.
The X-ray count rates are well consistent with the range
reported by \citet[][his Table~2]{schmitt-1997} for his sample of nearby ($< 13$ pc)
sun-like stars. Therefore, the fluxes do not require the presence of background
sources.
However, it cannot be ruled out.
Indeed, most of the galaxies would typically have X-ray fluxes 
at lower levels.
Comparable count rates \citep{anderson-et-al-2007} are primarily expected from 
active galactic nuclei (AGN).
The densities of AGNs at the X-ray flux level in question
are on the order of one per square degree \citep{anderson-et-al-2007},
rendering multiple alignments in the DUNES fields rather unlikely.

As for the optical/near-IR data, as demonstrated with HIP~73100, contaminating
background galaxies can, in principle, be identified via their optical counterparts.
We have not found such sources close to the stellar positions.
It is still possible though that the catalogs miss optically faint cold 
galaxies or other objects very close to the {\it Herschel} targets.

For all of the stars but HIP 29271, also raw VLA radio data are available from the 
NRAO Science Data Archive\footnote{https://archive.nrao.edu.}.
However, photometric results have only been published for HIP 49908,
with an unconstraining upper limit of $0.08\mJy$ at 3.6~cm \citep{guedel-1992}.
No extragalactic sources (e.g., AGNs) have been found here.

More VLA data can be found in the NRAO VLA Sky Survey \citep[NVSS,][]{condon-et-al-1998}
catalog. It lists a source in positional 
agreement with HIP~92043 and a tentative continuum flux of $2.7\mJy$ at 1.4~GHz (21~cm).
The average rms of $0.46\mJy$ \citep{condon-et-al-1998} for the NVSS suggests a 
SNR of 5--6. However, inspection of the noise
in the VLA field around HIP 92043 suggests the radio detection to be marginal at best
(K. Schreyer and M. Hoeft, pers. comm.).
Provided the radio source is real and taking into account
that a debris disk does not emit significantly at centimeter wavelengths,
both the far-IR excess and the radio counterpart could be caused by 
the dust emission and the synchrotron emission of a background AGN, respectively.
We invoke the expected relation between the far-IR and radio fluxes
of a typical AGN \citep[Eq.~(14) of][]{condon-et-al-1991}
to estimate that an AGN mimicking a disk with a far-IR fractional luminosity
$f_\mathrm{d} \approx 5\times 10^{-7}$ would be consistent with a radio flux 
$F_\mathrm{1.49~GHz} \approx 0.05\mJy$.
Thus the reported radio flux is off by a factor of 60,
inconsistent with the typical scatter width of only a factor of three observed for the 
far-IR-to-radio relation for AGNs \citep{condon-et-al-1991}.
Furthermore, the likelihood of an aleatory alignment of HIP~92043 with an AGN of a
plausible radio brightness is low. 
From the statistics presented in \citet{condon-et-al-1998}, there should be about 100 sources 
per square degree with $F_\mathrm{1.4~GHz}$ in the range from $1\mJy$ to $10\mJy$.
Within the synthesized FWHM of $45\arcsec$ for the NVSS, we would therefore expect 0.01 
such sources, on average.
HIP 92043 could hence be one of at most two expected sources in the DUNES sample 
aligned with AGNs of that magnitude.
In summary,
the marginality of the detection, inconsistency between the far-IR and radio 
fluxes, as well as low density of AGNs at the brightness level in question
all render the hypothesis of a coincidental alignment of HIP~92043 with an AGN
less likely than that of a classical debris disk plus a spurious radio signal.

\begin{figure}[htb!]
  \begin{center}
  \includegraphics[width=0.48\textwidth]{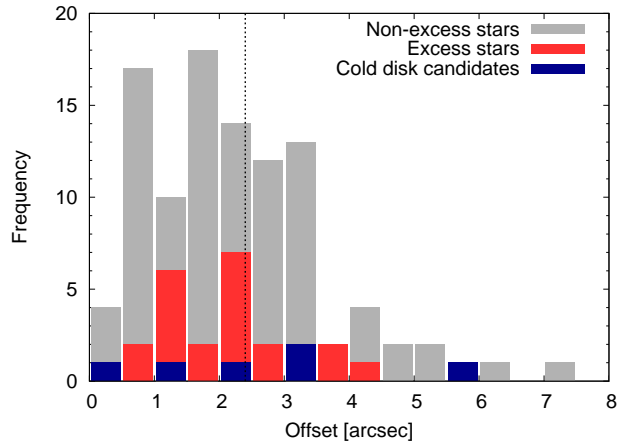}
  \end{center}
  \caption{
  Offsets of the peak of the $100\mum$ emission relative to the optical position
  of a star for all DUNES sources (grey bars),
  excess stars excluding cold disk candidates (red),
  and cold disk candidates (blue).
  One object (HIP~28442) with an anomalous offset of $10.5\arcsec$ was left out from the plot.
  The mean APE of {\it Herschel}/PACS is shown with a dashed vertical line.
  }
  \label{fig:offset}
\end{figure}

Apart from the direct search for possible background galaxies, we can
invoke statistical arguments:
\begin{itemize}
\item
For the cold disk candidates, the mean offset between the optical position of 
a star and the peak of the $100\mum$ emission
is as small as $2.6\arcsec$.
This is consistent with the {\it Herschel} 1-$\sigma$ absolute pointing error (APE)
of $2.4\arcsec$\footnote{This is the APE for the second period of {\it Herschel}
observations with scan maps, see\\
http://herschel.esac.esa.int/twiki/bin/view/Public/\\SummaryPointing.}.
The distribution of the offsets among the cold disk candidates is also consistent
with the distribution of offsets for other DUNES stars (Fig.~\ref{fig:offset}).
Five out of six have offsets $\le 3.2\arcsec$, and only HIP~171 has $5.5\arcsec$.
But even the latter is just a $\sim 2\sigma$ outlier, and thus
nothing extraordinary, especially for a binary.
It is known that the binarity may cause photometric shifts and decrease
the accuracy of the proper motion determination.
Note that there are five other non-excess sources in the DUNES sample with offsets 
$>5\arcsec$ \citep[][their Sect. 5.3]{eiroa-et-al-2013}.

\item
For three sources without an excess at $100\mum$ (HIP~171, HIP~29271, HIP~49908),
the measured flux at $100\mum$ is consistent with 
the photospheric prediction (mean deviation of 1.4~mJy,
well within the flux uncertainties),
strengthening the conclusion that the $100\mum$ emission indeed comes from the star.

\item
The mean offset between the peaks of the $160\mum$ emission and the $100\mum$ one is $1.8\arcsec$,
so that the probability that the $160\mum$ emission and the $100\mum$ one are associated
with each other is high.
\end{itemize}

%
%
%

These arguments are statistical and cannot be considered as evidence for the circumstellar
nature of the emission. Yet we can make probabilistic estimates.
Assume that, to be misinterpreted as a cold disk, a center of a
background galaxy should be located within a ``spot''
with a diameter equal to the FWHM of the PACS instrument at $160\mum$, i.e.~$12\arcsec$.
The area of a circle of that diameter is $0.0314$ square arcmin.
Next, we estimate the expected density of galaxies at a level of
$6\mJy < F(160\mum) < 13\mJy$,
which corresponds to the excess flux range of the cold disk candidates at that wavelength
(see Table~\ref{tab:six_data}).
We draw on the study by \citet{berta-et-al-2011} that derived the galaxy counts from 
several {\it Herschel} surveys, including GOODS and PEP. Their results are
also supported by the most recent analysis of the
{\it Herschel}/PACS data from the DEBRIS survey \citep{sibthorpe-et-al-2013},
with the caveat that the comparison has only been done at $100\mum$ and
for relatively bright sources.
Based on Figure~7 of \citet{berta-et-al-2011}, we have estimated the
density number of sources in the $6$--$13\mJy$ range to be about 5500
sources per square degree (1.53 sources per square arcmin).
This density is almost the same as the cumulative
density of all galaxies brighter than $6\mJy$ (the $dN/dS$ slope steepens
above $10\mJy$, see Fig.~7 in \citet{berta-et-al-2011}.
To estimate the likelihood of a coincidental alignment in the DUNES fields,
we assume that the source density is representative of any point of the sky,
so in a FWHM spot we will have $1.53 \times 0.0314 \approx 0.048$ sources,
implying the probability of a coincidental alignment of a given cold disk
candidate with a galaxy of $p = 4.8\%$.
We now consider the whole DUNES sample of $N = 133$ stars.
The mean number of false detections is simply
$N p = 6.4$, meaning that six or seven sources in the sample
are likely to be background objects.
However, as mentioned above, as many as seven of such sources seem
to have already been identified
\citep[see Sect. 5.3 of][]{eiroa-et-al-2013}.
The binomial probability that six more sources (namely, our cold disk
candidates) are all galaxies as well, is just $1.2\%$.
Nevertheless, the chance that at least one of our six is an unrelated object is $31\%$.
Conversely, there is a 69\% chance that all of them are true disks.

%
%
%
%

The above estimates are conservative for two reasons.
First, we used the density of galaxies brighter than $6\mJy$ at $160\mum$.
This density, and the resulting $p = 4.8\%$, were then applied to all six candidates,
although most of them are brigher than $6\mJy$.
Taking into account the excess flux of individual disks
(from $6.4\mJy$ for HIP~49908 to $12.6\mJy$ for HIP~92043)
would reduce the probabilities.
For example, the probability $p$ for $8\mJy$ sources such as HIP 171 and HIP 29271
is only $2.3\%$.
Second, DUNES observations were done at different depth,
with the on-source time ranging between $180\s$ and $1440\s$.
However, a $180\s$ observation would not be enough for a 3-$\sigma$ detection of
a $6\mJy$ source. Specifically, a 3-$\sigma$
detection would only be possible for $\ge 11.1\mJy$ sources at $180\s$,
$7.8\mJy$ at $360\s$, ..., $4.5\mJy$ at $1440\s$.
This means that the expected number of false detections in the whole DUNES sample,
which we computed as $N p(>6\mJy)$, where $N=133$ and
$p(>6\mJy)=4.8\%$, is again an overestimate.
If, for instance, we did not 
count the $180\s$ fields, or applied $p(>11.1\mJy)$ to them, the resulting  
number would be smaller.

More accurate probability estimates, involving the brightness of individual sources
and different integration times, would not make much sense, because of a number of
additional uncertainties.
The actual diameter of the ``alignment spot'' that goes into the estimate of $p$
may be smaller than the FWHM as assumed here.
Besides, the density of galaxies in the DUNES fields may differ somewhat
from those in GOODS and other cosmological surveys.
Even more importantly,
the density of galaxies at the brightness level in question varies
from one individual DUNES field to another~-- for instance, the field
around HIP~171 is cleaner than the one around HIP~92043.
We note, however, that the density adopted above agrees reasonably well
with the number of background sources at the brightness level of the cold
disk candidates and with a similar $100/160\mum$ color, which are seen in 
the $2^\prime \times 2^\prime$ fields around the positions of the six candidates.
A detailed study of the contamination within the DUNES fields
is in progress (del Burgo et al., in prep.).

We conclude that, most likely, our set of six candidates contains {\em both}
real disks and unrelated background galaxies.
We also emphasize that HIP~73100 and HIP~92043 almost certainly host true
debris disks, evident in the $70\mum$ excess, regardless of whether their cold
components are real.
Beyond that, we do not see any possibility to distinguish between the real cases
and ``false alarms'', and observational prospects to find final answers are
discussed in Sect.~5.
In what follows, we assume that some of the cold disk candidates do
represent true circumstellar disks. In the light of observational uncertainties
described above, detailed modeling of the individual objects in our set
would not be warranted. Therefore, the subsequent analysis seeks
possible qualitative explanation for the phenomenon. Conceivable scenarios
are then checked against numerical simulations.

\section{Grain Sizes and Composition}

\subsection{Blackbody grains}

We start with getting a handle on the dust temperature 
in the cold disks.
To this end, we have computed blackbody emission
by placing the emitting material at a distance $R_\mathrm{disk}$ from each star.
The results were then scaled vertically to match the excess fluxes
at $100\mum$ and $160\mum$.
The results are shown in Fig.~\ref{fig:six}; the blackbody temperature
as indicated in the panels ranges from $23\K$ (HIP~49908) to $38\K$ (HIP~92043).

HIP~109378 is nicely consistent with blackbody grains.
For HIP~73100 and HIP~92043, there is an excellent agreement at $100\mum$ and $160\mum$.
However, those stars reveal an $\sim 3\sigma$ excess at $70\mum$
(see notes to Tab.~\ref{tab:six_data}),
which may be indicative of an additional inner dust component.
HIP~49908 provides weak constraints,
as the excess was only found at $160\mum$.
Finally, for HIP~171 and HIP~29271 the rise of the excess flux from $100$ to $160\mum$
is steeper than blackbody.
However, as the deviation from the blackbody model is $< 2\sigma$,
this may not be genuine.
If it is, this would either imply ``subthermal'' dust
(colder than blackbody) or indicate background galaxies.
These two objects might particularly be interesting for ALMA followups, as
discussed in Sect.~5.

\subsection{Compact grains of pure materials}

The nearly blackbody temperatures of grains require them either to be large
or, if they are small, to have low absorption in the visible.
In this section, we investigate, which of these two options
appears more probable. More generally, we try to find out which kind
of material the cold disks should be composed of to reproduce the observed
thermal emission.

We have chosen four material compositions: 
astrosil \citep{draine-2003a,draine-2003b},
olivine \citep{fabian-et-al-2001},
crystalline ice at $-60\,^\circ$C \citep{warren-1984},
and amorphous ice \citep{li-greenberg-1998}.
This choice is motivated by the fact that
silicates are traditionally assumed as a reference material composition
in debris disk modeling, while ice-rich material can also be expected,
considering that all cold disks have large radii ($\sim 100\AU$).
For instance, \citet{lebreton-et-al-2012} demonstrated that
icy inclusions improve markedly the fits to the SED of a  bright, resolved
debris disk of HD~181327.
A direct evidence for ices comes from the solar system studies.
Indeed, surfaces of many transneptunian objects contain
significant amounts of ice
\citep{barucci-et-al-2011}.
Many large ones  have geometric albedos 
well in excess of $50\%$.
For instance, Haumea's albedo is $70$--$75\%$ \citep{lellouch-et-al-2012},
Makemake includes a bright terrain with an albedo of $\sim 80\%$
\citep{lim-et-al-2012},
Sedna and 2010~EK$_{139}$ have $32\%$ and $25\%$, respectively \citep{pal-et-al-2012}.
Some scattered EKBOs have albedos of up to $85\%$ \citep{santossanz-et-al-2012},
while cold classical EKBOs, despite their smaller sizes, still have an 
average albedo of $17\%$ \citep{vilenius-et-al-2012}.
It is natural to expect that dust released from the surfaces of such objects would
be highly reflective, too.

As for sizes, we selected two grain radii: $10\mum$ and $1\mm$.
The former choice is a proxy for a typical cross section-dominating size
expected from collisional models of debris disks
\citep[e.g.][]{wyatt-et-al-1999, wyatt-dent-2002, krivov-et-al-2006,thebault-augereau-2007}.
The latter choice is meant to show the emission of grains
with a size parameter (for the {\it Herschel} wavelengths) exceeding unity.
At this point we intentionally limit ourselves to single sizes,
in order to get a clearer understanding of the simulation results in Sect.~4 that
involve size distributions.

\begin{figure}[htb!]
  \begin{center}
  \includegraphics[width=0.48\textwidth]{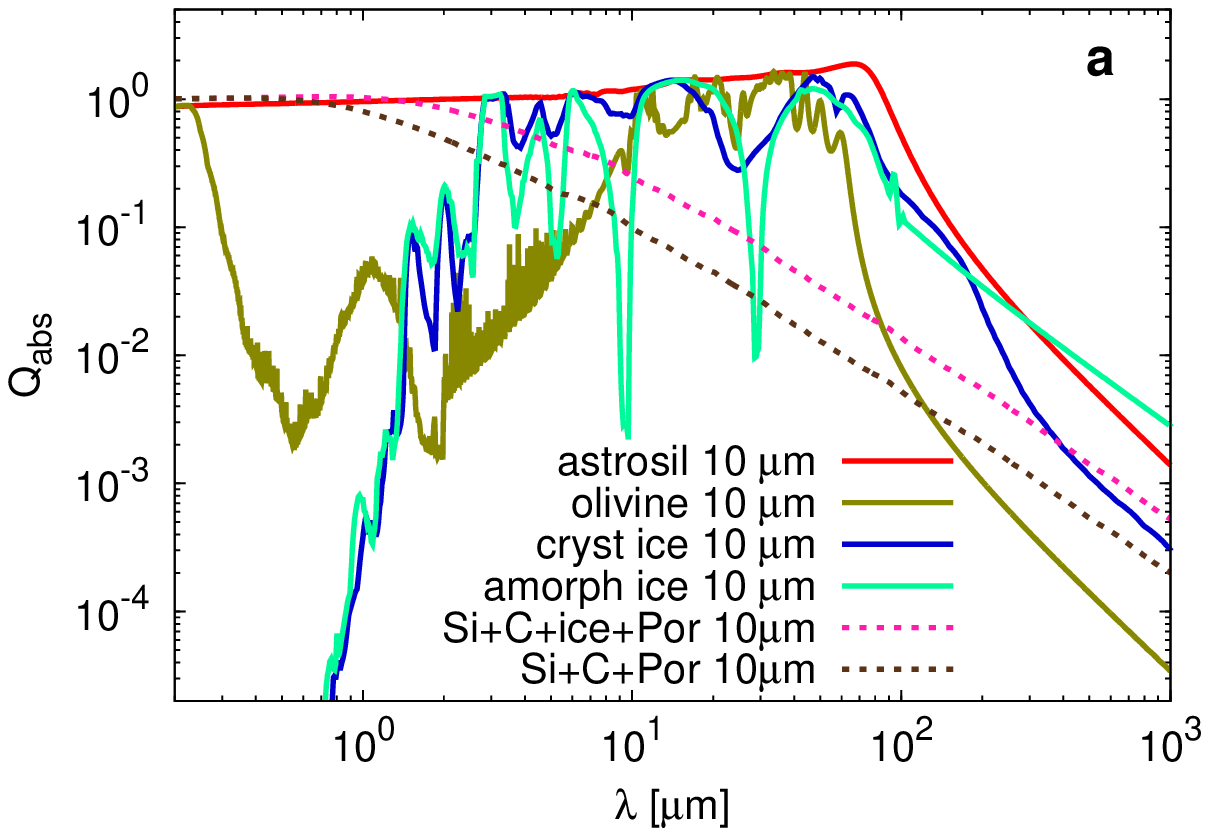}\\
  \includegraphics[width=0.48\textwidth]{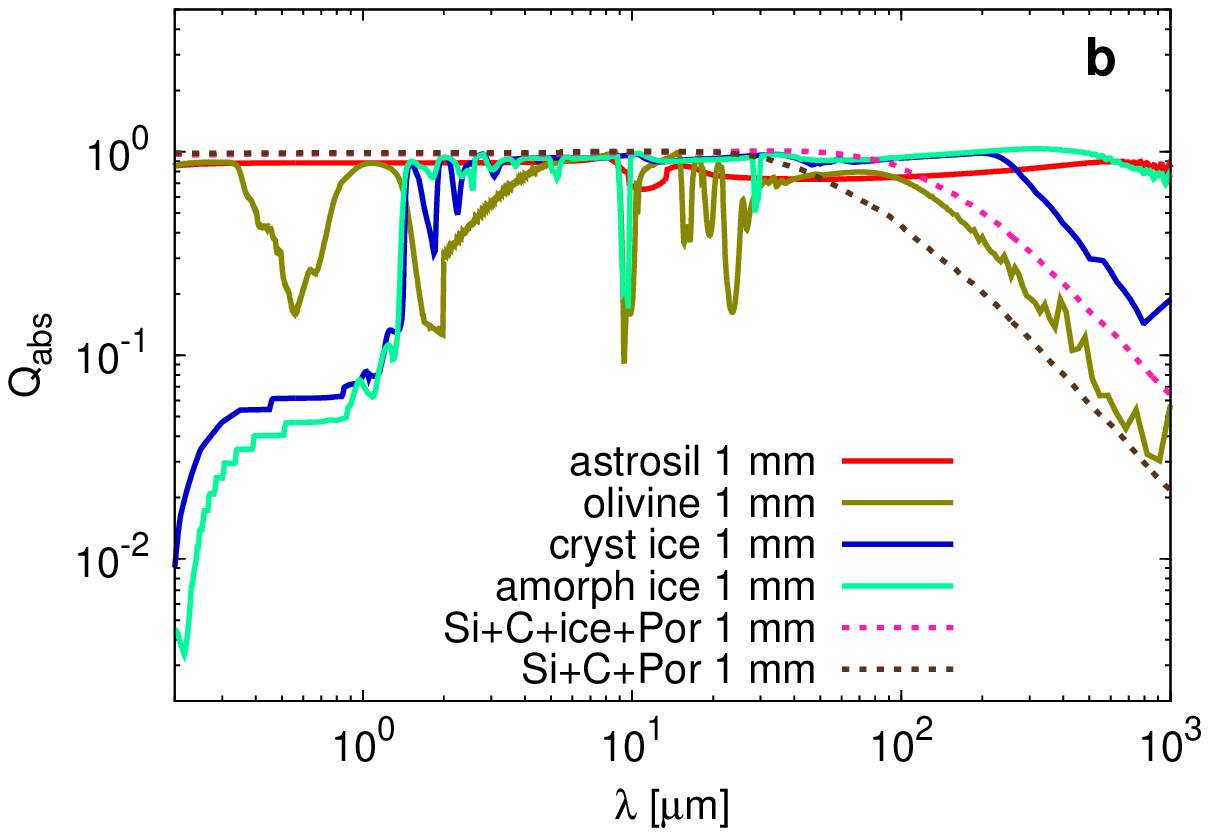}
  \end{center}
  \caption{Absorption efficiency
  for compact grains of single materials (solid lines) and
  for two advanced grain models (dashed lines).
  Grain radii: (a) $10\mum$ and (b) $1\mm$.
  }
  \label{fig:Qabs}
\end{figure}

The absorption efficiency $Q_\mathrm{abs}$ for all four materials and two sizes,
calculated with a standard Mie routine valid for homogeneous, 
compact spheres \citep{bohren-huffman-1983}, is plotted
in Fig.~\ref{fig:Qabs} with solid lines.
From these curves, we can expect icy grains of a given size at a given
distance from a star to be colder than silicate ones.
This is because icy grains, even mm-sized ones, have lower absorption
efficiencies in the visible (where the stellar flux peaks).
At the same time, the absorption efficiency at far-IR wavelengths (where
thermal emission of dust peaks), is comparably high for all materials,
providing efficient cooling.

\begin{figure*}[htb!]
  \begin{center}
  \includegraphics[width=0.48\textwidth]{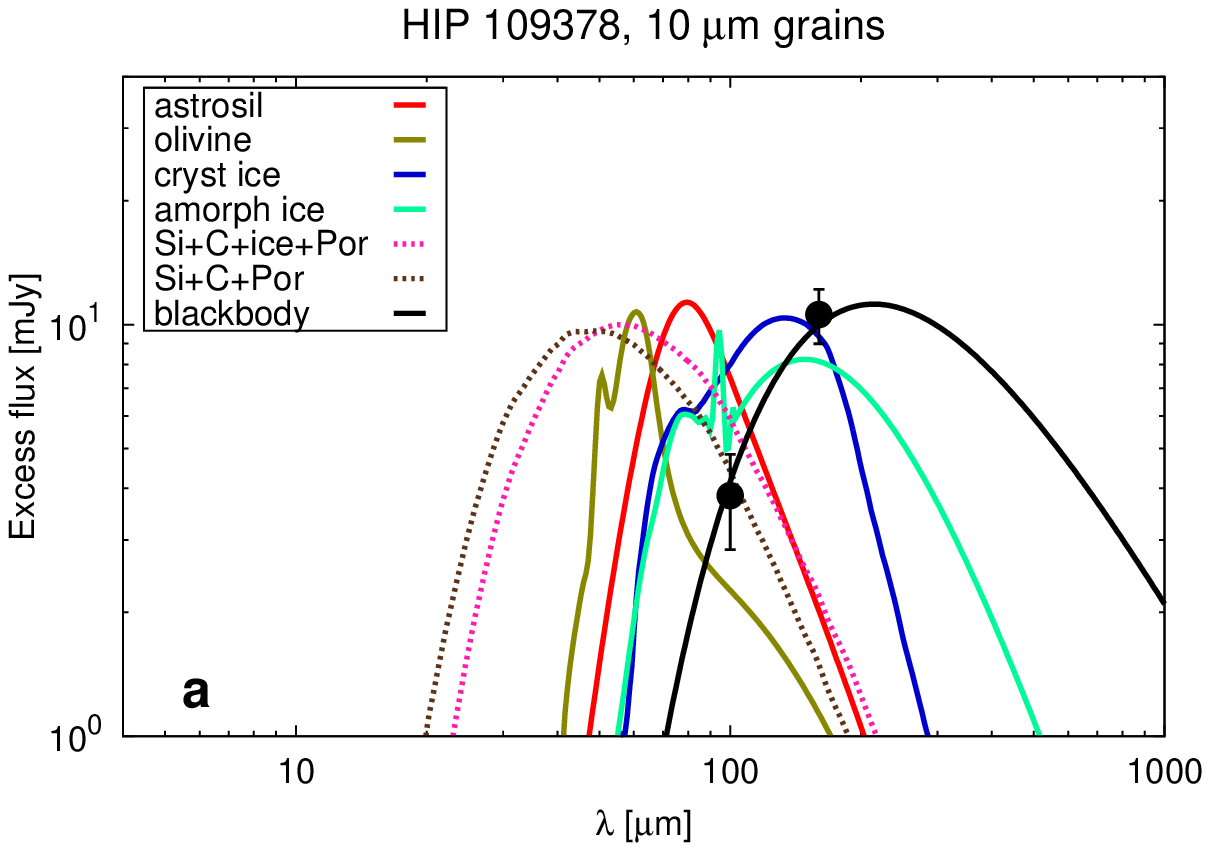}
  \includegraphics[width=0.48\textwidth]{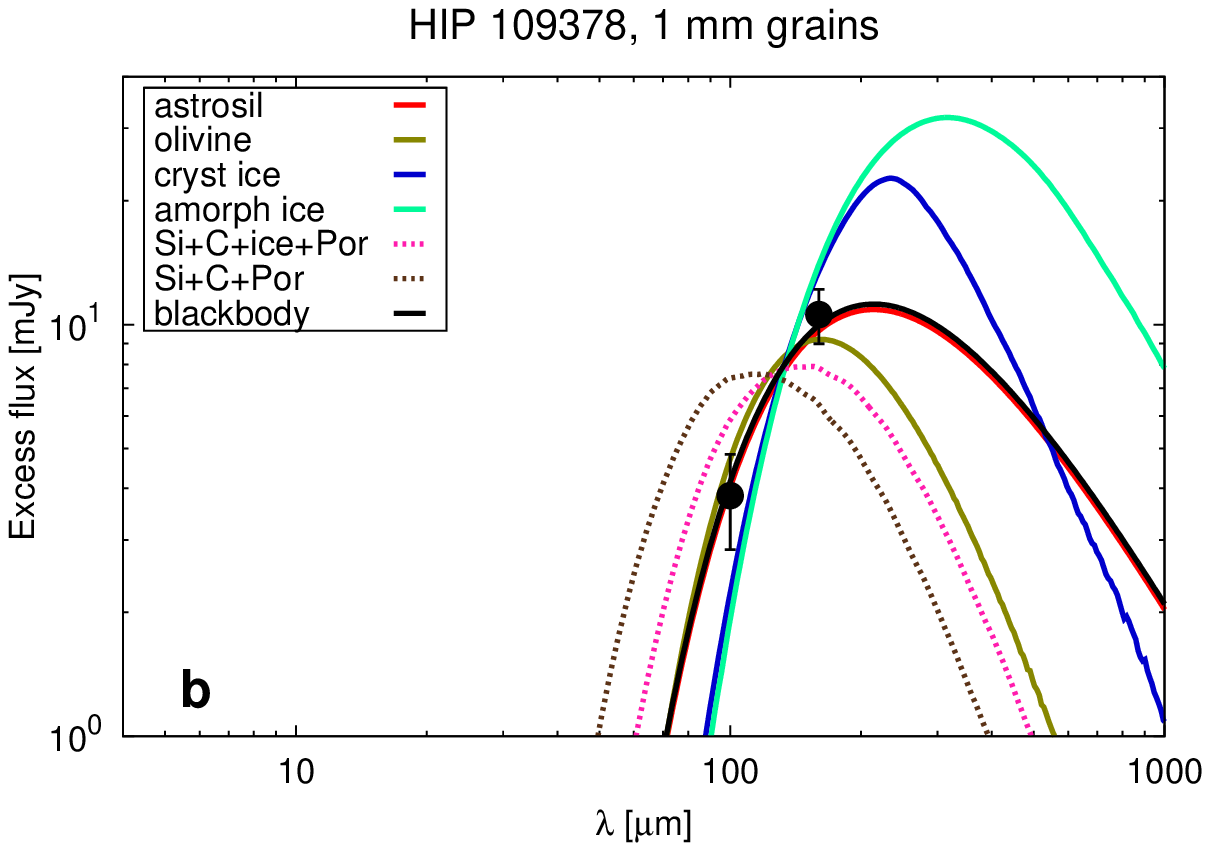}\\
  \includegraphics[width=0.48\textwidth]{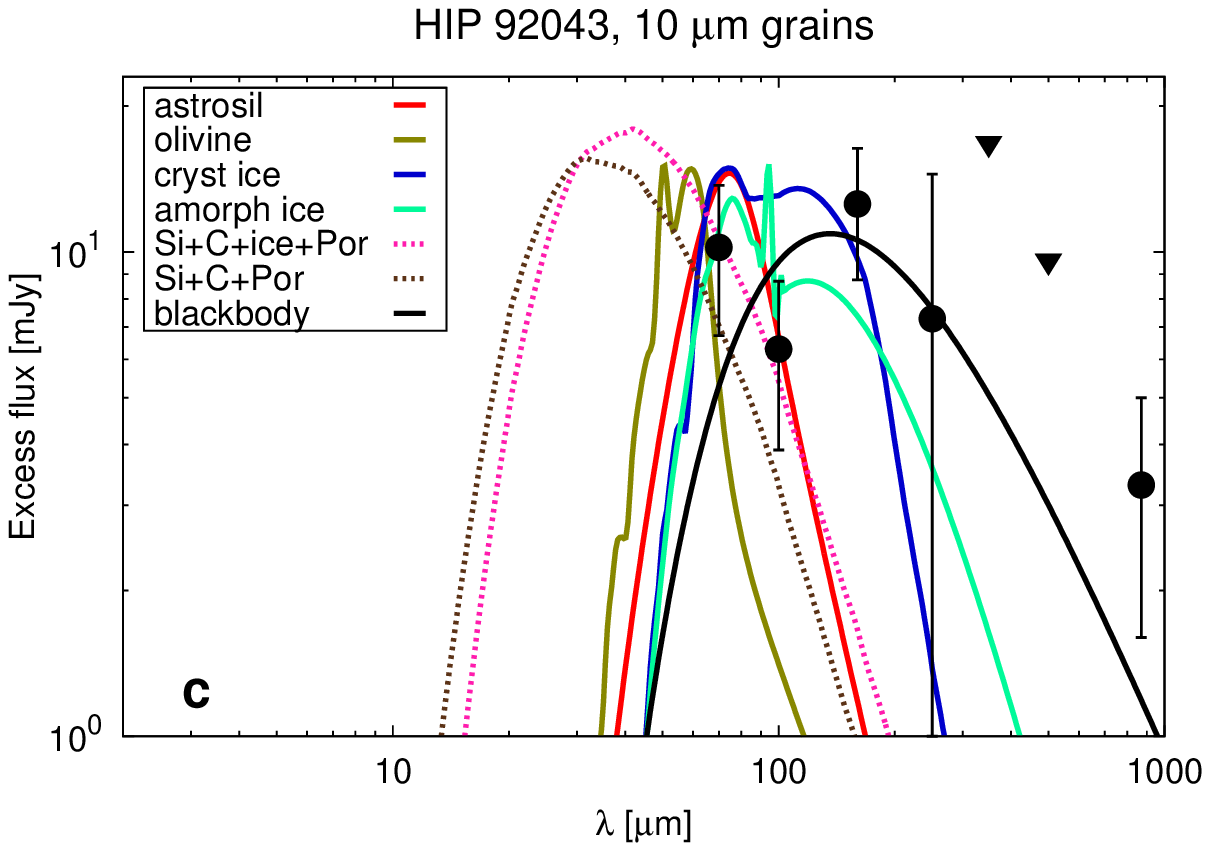}
  \includegraphics[width=0.48\textwidth]{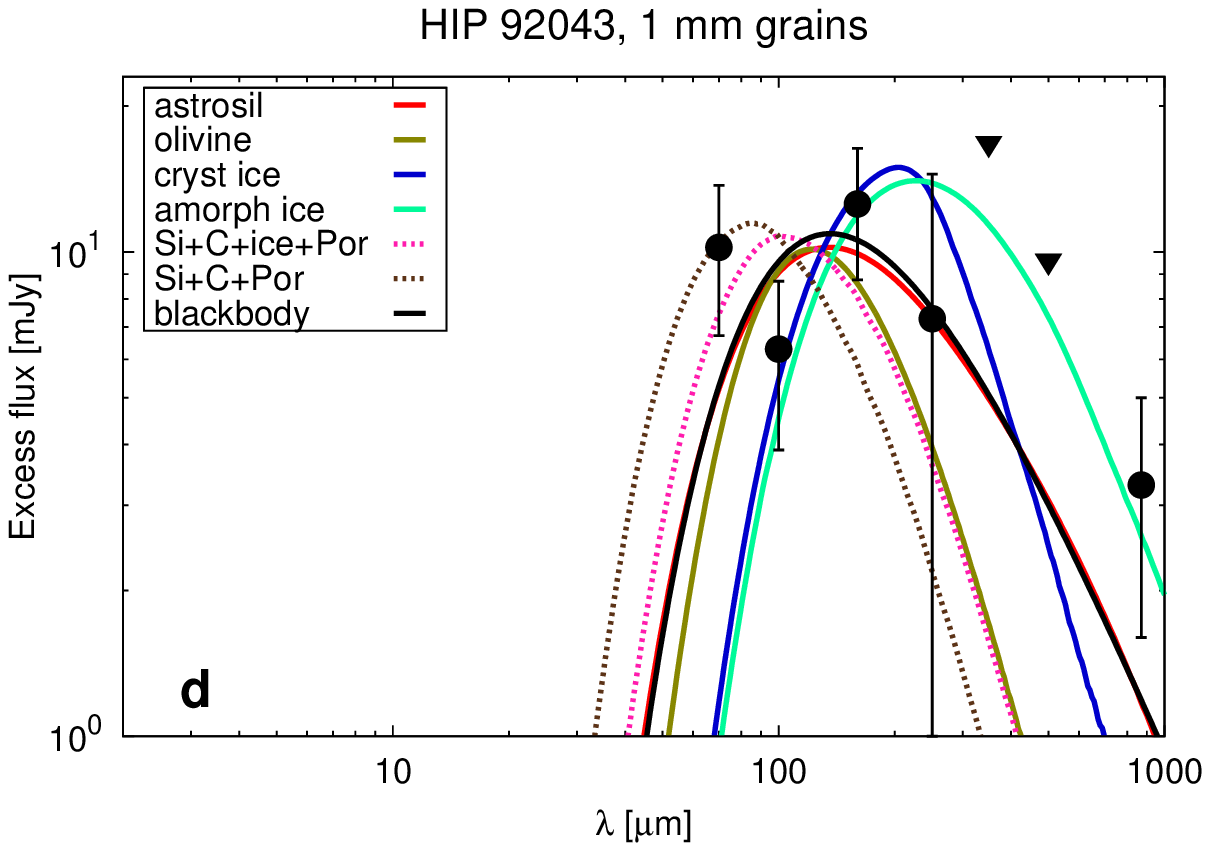}
  \end{center}
  \caption{
  Excess emission of  HIP~109378 (top) and HIP~92043 (bottom). 
  The stellar photosphere is subtracted.
  Filled circles with $1\sigma$ error bars are PACS/70, PACS/100, PACS/160, SPIRE/250,
  and LABOCA/870 excess fluxes.
  Upside-down triangles are the 1-$\sigma$ upper limits (SPIRE/350 and SPIRE/500) for HIP~92043.
  Colored lines are the same as in Fig.~\ref{fig:Qabs} and
  show which emission is expected from dust grains of
  $10\mum$ (left panels) and $1\mm$ in radius (right panels), made of
  several pure materials (solid lines) and mixtures (dashed).
  For comparison, blackbody emission is shown as a black thick curve.
  All curves are roughly scaled vertically to the
  level of the PACS/100 and /160 and, for HIP~92043, also SPIRE/250 points.
  }
  \label{fig:sizes-materials}
\end{figure*}

With these materials and sizes, we have calculated dust emission
for two selected candidates (Fig.~\ref{fig:sizes-materials}).
One is the ``clean'' case of HIP~109378, where 
there is a clear, point-like, excess emission at PACS/$100$ and PACS/$160\mum$
\citep{eiroa-et-al-2011}.
From the fact that the disk is unresolved the upper limit on the disk radius is
$\sim 130\AU$.
Another case is HIP~92043, which offers the best-sampled SED.
Here, excesses were found in the PACS/$70\mum$ band (and previously in MIPS/70),
in the PACS/$100$, PACS/$160$, and SPIRE/$250\mum$ bands.
The target was not detected in the $350$ and $500\mum$ SPIRE bands.
Finally, there was a marginal (slightly over $2\sigma$) detection by 
APEX/LABOCA at $870\mum$ (R. Liseau, pers.~comm.).
The emission at $70\mum$ and $100\mum$ is point-like,
but at $160\mum$ extended, and we interpret it as stemming from the disk.
Then, the deconvolved brightness profiles
suggest the disk radius of $\sim 130\AU$.
Since the excess fluxes at $70\mum$ and $160\mum$ are separated by
a lower excess flux at $100\mum$, and the images at $70\mum$ and $160\mum$
show different appearance of the emission (point-like vs extended),
it is likely that the $70\mum$ excess derives from an additional component,
an unresolved inner debris disk. If true, the potential cold disk should 
be associated mostly with the $160\mum$ emission.

Figure~\ref{fig:sizes-materials} overplots the expected thermal emission from 
fiducial disks of equal-sized grains with different optical properties,
placed at a distance
of $130\AU$ from a G0V (HIP~109378) and F6V (for HIP~92043) star
and assumed to be in thermal equilibrium.
For HIP~109378, a comparison of the modeled curves and the data points clearly 
demonstrates that grains smaller than $\sim 10\mum$~-- those that are predicted by
collisional models~-- yield emission that is much too warm to be consistent
with the observed fluxes, irrespective of the dust composition.
A rough agreement with the observations can only be achieved
for larger grains ($\ga 1\mm$).
Constraints on the material composition are difficult to place.
Similar conclusions can be drawn for two other sources, HIP~171 and HIP~29271,
not shown in the figure. 

For HIP~92043, the conclusions are less certain. However, a conclusion on
the prevalence of large grains can also be drawn here if, as argued above, the
$70\mum$ excess originates from an unresolved inner component, physically
separated from the bona-fide cold disk.
Again, constraints on the material composition are much weaker than those on sizes.
Although using amorphous ice provides a better agreement with the sub-mm point,
this should not be overinterpreted, as the LABOCA detection is only marginal.
The case of HIP~73100, not shown in the figure, is very similar to HIP~92043.

\subsection{Porous grains of material mixtures}

Real dust grains in various cosmic environments, including circumstellar
disks, are expected to be composed of material mixtures, to have some degree of
porosity and a complex morphology, for instance to acquire icy mantles
around silicate or organic refractory cores \citep[e.g.][]{preibisch-et-al-1993}.
Ice mantle growth and dust coagulation are very
common physical processes in the interstellar medium (ISM) where the
resulting fluffy, porous dust grains show emission enhancements
towards the far-IR \citep[][and references therein]{delburgo-et-al-2003}.
Multi-component fitting that includes core-mantle grains has also been
commonly done for protoplanetary disks, which represent the debris disk
progenitors \citep[e.g.][]{white-et-al-2000}.

To test how strong the effects of more advanced grain models could be,
and what impact on the observed emission they may have,
we tried two models from \citet{lebreton-et-al-2012}.
The first one provided the best fit to the SED of HD 181327.
It consists of $7\%$ volume fraction of ACAR carbon from \citet{zubko-et-al-1996},
$3.5\%$ astrosil from \citet{draine-2003},
$24.5\%$ amorphous water ice from \citet{li-greenberg-1998},
and 65\% vacuum to mimic porosity.
The bulk density is $0.553\g\cm^{-3}$.
Another model provided the ``coldest'' emission in the case
of HD~181327.
It has $3.3\%$ carbon, $1.7\%$ astrosil, no ice, but a
high porosity ($95\%$ of vacuum), and a low density,
$0.123\g\cm^{-3}$.
The refractive indices of both mixtures
were computed using the Bruggeman mixing rule \citep{bohren-huffman-1983},
and the Mie theory was used to calculate the absorption efficiencies.

The absorption efficiencies and SEDs for the same two cold disk candidates
are shown with dashed lines in Figs.~\ref{fig:Qabs} and~\ref{fig:sizes-materials},
respectively.
Indeed, the effects are seen to be strong, but~-- in contrast to HD~181327~--
go in the wrong direction: the SEDs for these two models are warmer, not colder,
than those discussed before.
This is because the porous grains absorb efficiently where the stellar flux peaks,
while these are poor emitters in the IR
(Fig.~\ref{fig:Qabs}).
As a consequence, the porous grains used here are hotter (at a same distance,
which is fixed), and the SED is shifted to the blue.

Nonetheless, this is not to say that realistic grain models are unable to make
the dust colder.
In fact, the opacities and temperatures of such grains can be both higher
and lower than those of compact mono-material ones, depending on the
chemical composition, sizes and number of constituent monomers, degree of porosity,
wavelength ratio between the maxima of the stellar and dust emission, and other
factors \citep[see, e.g.,][]{kimura-et-al-1997}. Also, the results of the
Mie calculations may \citep{mukai-et-al-1992} or may not \citep{stognienko-et-al-1995}
be accurate enough.
Finally, the mechanical properties (e.g., critical fragmentation energy)
of non-ideal grains are expected to be different from those of compact ones
\citep[e.g.,][]{love-et-al-1993,stewart-leinhardt-2009,shimaki-arakawa-2012}.
Therefore, for self-consistency, we would have to take porosity into account in the
collisional modeling performed in Sect.~4.
Taking into account the paucity of the observational data and the as yet unconfirmed
circumstellar nature of the ``cold emission'' of our candidates, we defer including
advanced grain models to future studies and do not consider them in the
rest of the paper. 

\section{Possible Scenarios}

Having concluded that the cold disks should be depleted in
small grains, in this section we seek a possible explanation for that.
We try to find a dynamical regime for a debris disk, which would simultaneously
provide the flux at an observed level and have a cross section peaking at larger sizes.

The simulations described below were performed with our collisional code ACE
\citep{krivov-et-al-2005,krivov-et-al-2006,krivov-et-al-2008,loehne-et-al-2011}
and thermal emission utility SEDUCE \citep{mueller-et-al-2009}.
These tools have been successful in reproducing SEDs and resolved images
of many other debris disks, ranging from the archetype debris disk around Vega
\citep{mueller-et-al-2009} to debris disks of the DUNES program
such as HD~207129 \citep{loehne-et-al-2011}.

The collisional code ACE numerically solves
the kinetic equation with various gain, loss, and transport mechanisms
to evolve a disk of solids in a broad range of sizes (from smallest dust grains
to planetesimals), orbiting a star under the combined action
of gravity, direct radiation pressure, and drag forces and experiencing
non-elastic collisions.
The code implements a three-dimensional model with masses, periastron
distances, and eccentricities as phase-space variables. It assumes the
disk to be rotationally symmetric and averages over the inclinations
of the constituent particles within the semi-opening angle of the disk.
Collision outcomes are simulated as follows.
The mechanical strength of the disk material is described by the
critical energy for fragmentation and dispersal:
\be \label{QD}
   Q_D^\ast (s)=
   A_s \left( s/1\m \right)^{b_s}
   +
   A_g \left( s/1\km \right)^{b_g}.
\ee
where $s$ is the radius of the target.
Unless otherwise stated, we choose
$A_s = 1.0 \times 10^6\erg/\mathrm{g}$, $b_s = -0.37$,
$A_g = 2.0 \times 10^6\erg/\mathrm{g}$, and $b_g = 1.38$,
which is close to the values used by many authors
\citep[see, e.g.,][]{davis-et-al-1985,holsapple-1994,paolicchi-et-al-1996,%
durda-dermott-1997,durda-et-al-1998,benz-asphaug-1999,kenyon-bromley-2004c}.
For each collision, ACE first checks if the impact energy,
$E_\mathrm{imp}(m_\mathrm{p},m_\mathrm{t})$, exceeds the critical one,  
$E_\mathrm{crit}(m_\mathrm{p},m_\mathrm{t}) \equiv
(m_\mathrm{p} + m_\mathrm{t}) \times Q_D^\ast(m_\mathrm{p}+m_\mathrm{t})
$,
where $m_\mathrm{p}$ and $m_\mathrm{t}$ are the masses of a projectile and target, respectively.
If it does, the collision is treated as disruptive.
If not, further checks are being done to determine whether the collision
is cratering (the projectile is disrupted, but the target is only cratered),
bouncing (both impactors are cratered),
or sticking (the two impactors merge into one).

For the purposes of this study, we made a number of further improvements to ACE.
All of these do not affect the modeling results as long as the relative
velocities exceed a few tens of $\m\s^{-1}$, but they provide a more accurate treatment
at lower velocities. First, the simulations are now being done
over a logarithmically spaced (instead of a linearly spaced) eccentricity grid.
Second, we have changed the prescriptions for the outcomes of
grain-grain collisions in the cratering, bouncing, and sticking regimes.
These are implemented in a model that approximately
conforms to a semi-empirical model of \citet{guettler-et-al-2010}.

\subsection{Disks in the transport-dominated regime}

One possibility to reduce the proportion of small grains is to assume that
the disks in question are transport- rather than collision-dominated
\citep{krivov-et-al-2000b,wyatt-2005},
meaning that small grains are displaced inward from their birth location
by Poynting-Robertson (P-R) drag before they get lost to collisions.
This can indeed be expected, since this regime is achieved at
normal optical depths lower than roughly $v_K/c$,
where $v_K$ is the local Keplerian speed \citep{kuchner-stark-2010},
which is consistent with low $f_d$ of the cold disks.
In that case,  the dominant size of the grain cross section
shifts to larger values
\citep{vitense-et-al-2010,wyatt-et-al-2011}, which is exactly what is needed.
However, it has to be checked whether our disks are tenuous
enough to fall into this regime.
Besides, even if they are, an additional difficulty may
arise, as smaller grains are not eliminated from the system.
Instead they drift inward, and heat up very efficiently, for them
being small and close to the star.
As a result, they could produce warm emission,
making the SEDs inconsistent with those of the cold disks.

To check this scenario, we ran ACE with
a setup that matches the conditions of the HIP~109378 disk.
The material was initially placed into a narrow ring
with a half-width of $10\AU$ centered at $130\AU$.
We assumed the disks to be composed initially of
planetesimals up to $s_\mathrm{max} \approx 1\m$ in radius with a differential size distribution
slope of 3.7
(this slope is expected for collisional equilibrium in the strength regime,
\citeauthor{o'brien-greenberg-2003} \citeyear{o'brien-greenberg-2003}).
The minimum size was set to $s_\mathrm{min} = 3\mum$, but the assumed $s_\mathrm{min}$ does not affect
the final results because the system immediately finds the equilibrium distribution
of grains at such sizes.
We tried different initial total disk masses: $0.1 M_\oplus$,
$10^{-3} M_\oplus$, and $10^{-5} M_\oplus$.
Henceforth we refer to these three as ``high-mass'', ``mid-mass'', and ``low-mass''
disks, respectively.
Note that the results of these simulations are almost independent of the
maximum size $s_\mathrm{max}$ chosen.
It only has to be large enough for the largest bodies not
to be involved in the collisional cascade by the time when the distribution
of material at dust sizes reaches a quasi-steady state \citep{loehne-et-al-2007},
and $1\m$ suffices for the disks considered here.
Choosing larger $s_\mathrm{max}$ would only slow down the simulations.
The results obtained here can readily be applied to larger $s_\mathrm{max}$,
except that an up-scaling of the disk mass is required.
For instance, if bodies as large as $100\km$ are present, the total
mass of the high-mass disk will be about $30 M_\oplus$ instead of $0.1 M_\oplus$.
The planetesimal disk was assumed to be moderately stirred to eccentricities $e \sim 0.1$ and
inclinations $I \sim 0.05$ (energy equipartition).
We took an homogeneous mixture of astrosilicate
\citep{draine-2003a,draine-2003b}
and amorphous water ice \citep{li-greenberg-1998}
in equal volume fractions ($\rho_\mathrm{d} = 2.35\g\cm^{-3}$).
The optical constants of the mixture were calculated by effective medium theory
with the Bruggeman mixing rule.
The mechanical strength of the disk material was set as described above.

We evolved all three disks until
a quasi-steady state \citep{loehne-et-al-2007} was reached:
after $\sim 0.5 \Myr$ for the high-mass disk, 
$\sim 50 \Myr$ for the mid-mass disk, 
and $\sim 5 \Gyr$ for the low-mass disk.
This reflects the property that the collisional timescales approximately go as
a reciprocal of the disk mass \citep{krivov-et-al-2008}.
The timescales listed above should not be misinterpreted as physical
time of collisional evolution of the disks.
Instead, these are merely the ``relaxation''
times for the fiducial collision-driven systems.

\begin{figure}[htb!]
  \begin{center}
  \includegraphics[width=0.48\textwidth]{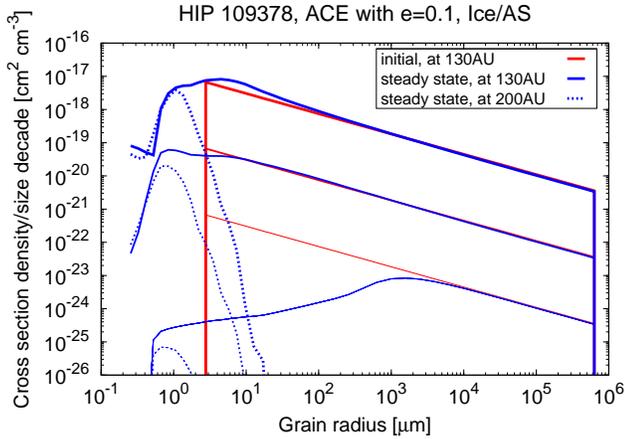}
  \end{center}
  \caption{
  Simulated evolution of the size distribution of  ``classical'' debris disks
  in an attempt to match the conditions of HIP~109378.
  Thick, medium, and thin lines: disks with initial total masses of
  $0.1 M_\oplus$, $10^{-3} M_\oplus$, and $10^{-5} M_\oplus$ in the bodies
  smaller than $1\m$, respectively.
  Red solid lines: assumed initial size distributions,
  blue solid: final distributions at $130\AU$,
  blue dashed: final distributions at $200\AU$.
  }
  \label{fig:transport_size}
\end{figure}

\begin{figure}[htb!]
  \begin{center}
  \includegraphics[width=0.48\textwidth]{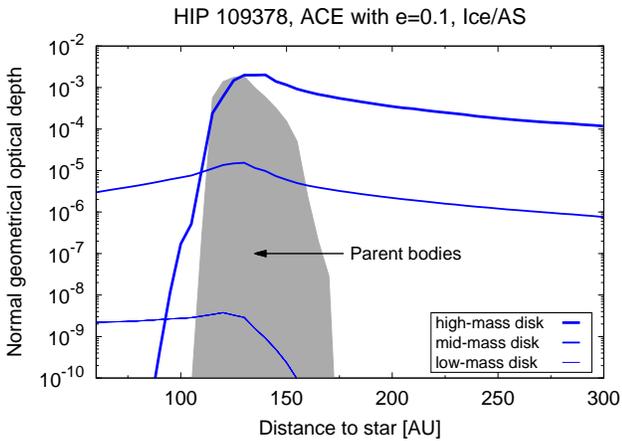}
  \end{center}
  \caption{
  Radial profiles of the same disks as in Fig.~\ref{fig:transport_size}.
  The filled grey area shows the distribution of large,  $1\mm$-sized grains
  (which we scaled vertically to the height of the other curves)
  to illustrate the position of the ``parent ring''.
  }
  \label{fig:transport_tau}
\end{figure}

The simulation results are presented in
Figs.~\ref{fig:transport_size} and~\ref{fig:transport_tau}
that plot the size distributions and the radial profiles of the optical
depth, respectively.
Obviously, the high-mass disk is collision-dominated, whereas
the low-mass one is transport-dominated.
In the former case, the size distribution peaks at grains as small as
several $\mum$, and the radial profile extends inward from the
parent ring position only moderately.
In the latter case, we observe the expected significant shift in the
size of grains that dominate the cross section to several hundreds of $\mum$,
and the radial profile reveals that transport efficiently 
fills the inner void of the disk with dust.
The transition from the collision- to the transport-dominated regime is
found to occur around the dustiness level of the mid-mass disk.
The size distribution in that disk still resembles that of the
high-mass disk, but the inner gap in the disk is essentially filled
with material, similar to the low-mass disk case.

\begin{figure}[htb!]
  \begin{center}
  \includegraphics[width=0.48\textwidth]{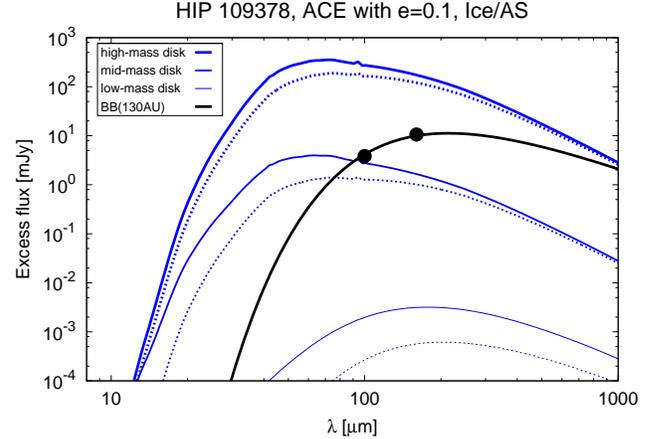}
  \end{center}
  \caption{
  SEDs of the disks presented in Figs.~\ref{fig:transport_size} and~\ref{fig:transport_tau}
  (blue solid lines).
  The contributions from the ``parent ring''
  (more exactly, from region between $120\AU$ and $140\AU$)
  are shown with dashed lines.
  The ``blackbody'' SED (thick solid line) and the observed excess emission of 
  HIP~109378
  (symbols; error bars are smaller than the data points) are plotted for comparison and 
   are the same as in Fig.~\ref{fig:sizes-materials} (top).
  }
  \label{fig:transport_sed}
\end{figure}

These results bring up the major question:
is the optical depth of our cold disk
candidates, exemplified by HIP~109378 here, low enough to put them
into the transport-dominated regime?
To answer this, we used SEDUCE to calculate the SEDs of the three modeled disks and
compared them with the observed emission level in
Fig.~\ref{fig:transport_sed} (solid lines).
The answer is clearly negative.
The optical depth of the disk of HIP~109378 (and the other cold
disk candidates) lies between the high- and mid-mass cases.
As a result, the dominant grains are still in the $\mum$ range,
and the modeled SEDs of both the high-mass and mid-mass disk
are too ``warm''.
They would imply that the excess emission at $70\mum$ 
is stronger than at $160\mum$ (and that the excess emission
at $70\mum$ is extended).
This is not observed, even for HIP~92043 and HIP~73100
that do reveal a $70\mum$ excess.

Furthermore, even if the disk had an optical depth low enough
to match the low-mass regime, this  solution would have to be
ruled out because of the unavoidable contribution of the warmer small
grains, drifting to the inner regions from the parent ring.
This is clearly apparent from Fig.~\ref{fig:transport_sed}.
Dashed lines plot the contributions
to the overall emission made by the material in the parent ring
region. Judging by the difference between the solid and dashed curves,
we conclude that, indeed, the dust inside the ring leads to a noticeable
increase of the fluxes and slightly shifts the peak of the SEDs towards
shorter wavelengths. (Note that the contribution of the ``halo'' grains
outside the ring is negligible. We have checked that, for instance,
placing the outermost distance at $300\AU$ instead of $200\AU$ leaves
the SEDs almost unchanged.)

\subsection{Disks with low dynamical excitation}

Another possibility is to assume that dust-producing planetesimals
have a low dynamical excitation which, however, is still high enough for
collisions to be mostly destructive.
In this case, low collision velocities between large grains, insusceptible
to radiation pressure, would create an imbalance between the rates at which small
grains are produced (low) and destroyed (high).
As a result, the disk would be devoid of small particles \citep{thebault-wu-2008}.

This scenario has been confirmed previously with detailed ACE simulations
by \citet{loehne-et-al-2011},
who modeled the {\it Herschel}/DUNES disk around HD~207129
\citep{marshall-et-al-2011}.
Similar to our cold disk candidates, it has
a solar-type central star and a radius of $\ga 100\AU$,
but its fractional luminosity is
appreciably higher and the emission is warmer.
\citet{loehne-et-al-2011} have shown that a low dynamical excitation
with  $e \sim I \sim 0.03$ shifts the grain size at which the size 
distribution peaks from $\approx 0.5\mum$ to $\approx 3$--$4\mum$.
However, to be able to explain the cold disks, the effect needs to
be stronger, and the dynamical excitation lower.
Indeed, for the size distribution to peak at $\sim 100\mum$
in the \citet{thebault-wu-2008} scenario, we would need $e \sim 0.001$.
Note that \citet{thebault-wu-2008} made their simulations for A-type stars,
for which the blowout size is larger than for FGK stars considered here,
which would necessitate even lower eccentricities.
The eccentricities $e \la 0.001$
would correspond to relative velocities of $\la 3\m\s^{-1}$.
Numerous laboratory experiments and microphysics simulations
\citep[see][for a review]{blum-wurm-2008}
tell us that at this level, the collisions are not necessarily destructive
and the scenario of \citet{thebault-wu-2008} in its proposed form
may no longer be applicable.

To check more quantitatively what exactly happens at excitation levels
lower than those previously simulated with ACE ($e \sim I \sim 0.03$),
and to what extent the effects predicted
by \citet{thebault-wu-2008} come into play,
we now explore the $e \sim 0.01$ regime.
We ran ACE for the 
disk around HIP~109378, and assumed a planetesimal ring centered at $130\AU$, initially composed of
large grains with initial radii between $1\mm$ and $\approx 1\m$
(with a differential size distribution slope of 3.7).
Note that the maximum size was set to $\approx 1\m$ arbitrarily;
we could have taken any value below $\sim 30\km$
(see estimates in the end of Sect.~4.4),
for which the largest planetesimals do not yet stir the disk to higher values of
eccentricity and inclination than the ones we assumed in this simulation.
The same applies to the minimum size: we made a separate test run with
$s_\mathrm{min} = 1\mum$ instead of $1\mm$ and did not see any appreciable difference.
As in Sect.~4.1, we took a homogeneous mixture of astrosilicate and water ice
in equal volume fractions.
The choice of the initial total disk mass is trickier, since we have to
come up with the fluxes at the observed level, but we do not know in
advance how the evolved size distribution will look like.
With a few ``guess-and-try'' attempts, the ``right'' disk mass
was finally found to be $0.02 M_\oplus$.
We used uniformly distributed eccentricities
between 0.005 and 0.015 and inclinations of $\le 0.01$,
which would correspond to relative velocities of $\sim 30\m\s^{-1}$.
In this velocity regime, in contrast to that considered in Sect.~4.1,
we enter the conditions probed by numerous direct laboratory impact 
experiments.
Accordingly,
we used a flat (index $q=2.0$) size distribution of fragments,
as suggested by experiments \citep{guettler-et-al-2010}.
The other parameters, including $Q_D^\star$, were the same as in the previous runs
(see  Sect.~4.1).

\begin{figure}[htb!]
  \begin{center}
  \includegraphics[width=0.48\textwidth]{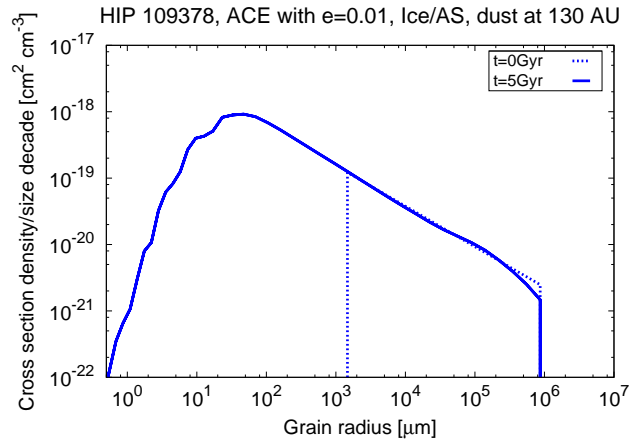}
  \end{center}
  \caption{
  Simulated evolution of the size distribution in a disk with $e\sim 0.01$
  macroscopic grains under the conditions of HIP~109378.
  Dashed: assumed initial distribution,
  solid: distribution after $5\Gyr$ of collisional evolution.
  }
  \label{fig:tbo-wu_size}
\end{figure}

\begin{figure}[htb!]
  \begin{center}
  \includegraphics[width=0.48\textwidth]{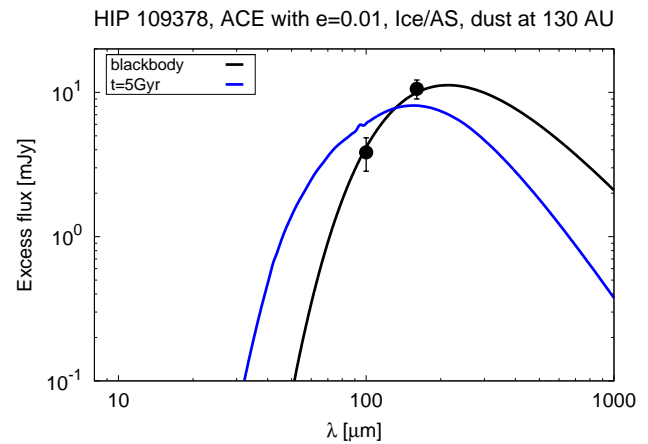}
  \end{center}
  \caption{
  SEDs of the same disk as in Fig.~\ref{fig:model_size} (blue line).
  The ``blackbody'' SED (black line) and the observed excess emission of HIP~109378
  (symbols with error bars which are smaller than the symbol size)
  are shown for comparison and are the same as in Fig.~\ref{fig:sizes-materials}a,b.
  }
  \label{fig:tbo-wu_sed}
\end{figure}

The evolution of the size distribution after $5\Gyr$ is shown in 
Fig.~\ref{fig:tbo-wu_size}.
These results demonstrate that collisional production of small grains
is still quite efficient.
Although there occurs a shift of the dominating
sizes towards large values, as predicted by \citet{thebault-wu-2008},
the effect is not strong enough.
The dominating grains are only a few tens of micrometers in radius.
The resulting SED which is shown in Fig.~\ref{fig:tbo-wu_sed},
although coming closer to the available photometry points of HIP~109378,
still appears too ``warm''.

\subsection{Unstirred disks with  ``jostling'' collisions}

We now explore the regime of {\em very} low dynamical excitations,
and {\em very} low collisional velocities.
We start with simple analytic considerations.
Since we are now interested in a disk at a very
low dynamical excitation level, one may be wondering whether collisions
in such a disk are frequent enough to play any role at all.
Assuming grains in almost circular, non-inclined orbits, one might
expect to get a system with very long collisional timescales, perhaps
exceeding the system's age.
However, this is not true.
For a narrow disk with a radius $r$ composed of equal-sized grains,
the collisional timescale is of the order of
\citep{backman-paresce-1993,artymowicz-clampin-1997, wyatt-et-al-1999,wyatt-2005}
\be
 T_\mathrm{coll} = {P \over 2\pi \tau_\perp} ,
\label{eq:Tcoll}
\ee 
where $P$ is the orbital period at $r$ and
$\tau_\perp$
is the normal geometrical optical depth of the disk at that distance.
Interestingly, Eq.~(\ref{eq:Tcoll}) does not depend on eccentricities
and inclinations. This is because, in the first order in $e$ and $I$,
$T_\mathrm{coll}$ is proportional to the ``effective interaction volume of particles''
and inversely proportional to their relative velocity. Both, in turn, are
proportional to the eccentricities and inclinations, so these cancel out
\citep[see, e.g.][their Eqs.~15--18]{krivov-et-al-2006b}.

For each disk, the dust luminosity and the position
of the SED maximum are known, at least roughly.
These two quantities uniquely determine
both the total cross section of material and the characteristic
disk radius $r$ which, via (Eq.~\ref{eq:Tcoll}), immediately
fix $T_\mathrm{coll}$. 
Although real disks must have a distribution of grain sizes
(making $T_\mathrm{coll}$ size-dependent) and may be radially extended
(making $T_\mathrm{coll}$ larger, since the same amount of emission
would be reached at a lower $\tau_\perp$), Eq.~(\ref{eq:Tcoll})
allows a ball-park estimate. For a disk around
a solar-type star with an optical depth of $\tau_\perp = 10^{-6}$
and a radius of $r = 100\AU$, we have $T_\mathrm{coll} \sim 200\Myr$.
Since the host stars of the cold disks have ages of Gyrs
(see Tab.~\ref{tab:six_parms}),
we conclude that the collisions cannot be ignored.
This conclusion will be directly confirmed by collisional simulations
described below.

Since the system will not be collisionless even at low dynamical
excitations, we can only try to find a range of relative velocities
that would allow macroscopic primordial grains ($s \ga 1\mm$) to survive,
at the same time not producing too many ``unwanted'' small grains.
The velocities  should not be too large, otherwise collisions
will be too destructive, creating too many small fragments.
They should not be too small either,
because collisions at very low velocities
will be 100\% sticking, leading to a rapid loss of the net cross section of
the material.
In fact, the results of laboratory experiments and numerical simulations
uncover a very complex view 
\citep[see][and references therein]{blum-wurm-2008}.
They show that the collisional outcome depends on impact velocity and impact angle,
masses, materials, porosities, and hardnesses of projectile and target,
radius of curvature of the target surface, morphology; size ratios of the impactors,
and other factors.
The bouncing regime, which would be the most favorable for the cold disk scenario to work,
typically occurs at sizes $1$--$10\cm$, or velocities from a few $\cm\s^{-1}$
to a few $\m\s^{-1}$.
Note, however, that the behavior depends sensitively on
the grain morphology \citep{poppe-et-al-2000}.
As an example, spherical particles possess a rather well-defined threshold velocity
around $1\m\s^{-1}$, below which they always stick and above which they never do.
However, irregularly shaped dust particles may stick even at impact velocities of several
tens of meters per second in ``hit-the-wall'' collisions.

In what follows, we check whether the desired regime can be reached
at relative velocities of a few $\m\s^{-1}$.
This level of dynamical excitation 
(relative velocities of a few meters per second, or
$e \sim I \sim 0.001$ for our cold disks)
roughly agrees with the one that can be expected
for the solids at the beginning of their collisional evolution, i.e.,
shortly after the dispersal of the primordial gas.
Indeed, that level should
be largely determined by turbulent velocities of the gas phase.
At radii of $100\AU$, one expects random velocities in the $\m\s^{-1}$
range for smallest grains
that were strongly coupled with gas at the protoplanetary (T~Tau) phase
and somewhat higher ones ($\sim 0.1$ of the sound speed) for weakly coupled ones
(R. Nelson, pers. comm.).

To see whether the amount of small dust can be suppressed to 
a sufficient extent in this very low velocity scenario,
we again used ACE.
The setup was the same as in Sect. 4.2, except that
we decreased the dynamical excitation by one order of magnitude.
Specifically, we assumed uniformly distributed eccentricities
between $0.0005$ and $0.0015$ and inclinations of $\le 0.001$,
resulting in relative velocities of $\sim 3\m\s^{-1}$.
Besides, we made three runs with dissimilar material strength
and grain stickiness.
In one run (``ref''), we 
use the standard values of $Q_D^\star$ (see text after Eq.~\ref{QD}).
In a separate (``sticky'') run we kept the same $Q_D^\star$,
but assumed that in every cratering or bouncing collision,
a fraction of cratered mass equal to $1 - E_\mathrm{imp}/E_\mathrm{crit}(m_\mathrm{p},m_\mathrm{t})$
remains stuck to the target instead of being ejected to space.
Such an increased  ``stickiness''
was recently found experimentally for low-temperature ice
by \citet{windmark-et-al-2012}.
Finally, in the ``weak sticky'' run we assumed the same prescription
of stickiness, but reduced $Q_D^\star$
by two orders of magnitude, as reported by \citet{beitz-et-al-2011}.
The total masses of the three disks, needed to produce the dust emission at the observed 
level, were again determined with an iterative ``guess-and-try''
procedure to be $0.07M_\oplus$ (``ref''), $0.25M_\oplus$ (``sticky''),
and $0.8M_\oplus$ (``weak sticky'').

\begin{figure}[htb!]
  \begin{center}
  \includegraphics[width=0.48\textwidth]{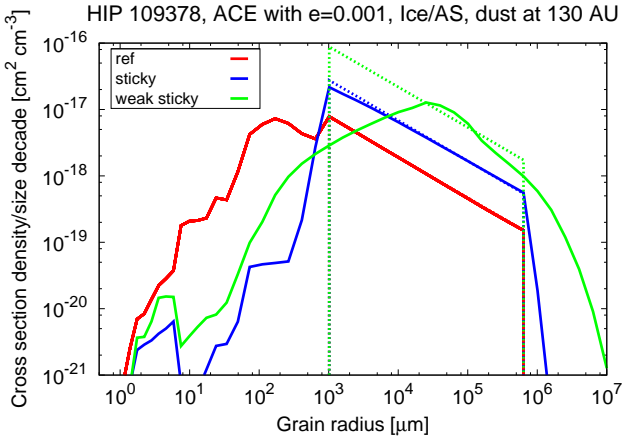}
  \end{center}
  \caption{
  Simulated evolution of the size distribution of a dynamically cold belt of 
  macroscopic grains under the conditions of HIP~109378.
  Dashed lines: assumed initial distributions,
  solid: distributions after $5\Gyr$ of collisional evolution.
  Red, blue, and green lines correspond to the ``ref'', ``weak'', and ``weak sticky''
  runs as described in the text.
  }
  \label{fig:model_size}
\end{figure}

\begin{figure}[htb!]
  \begin{center}
  \includegraphics[width=0.48\textwidth]{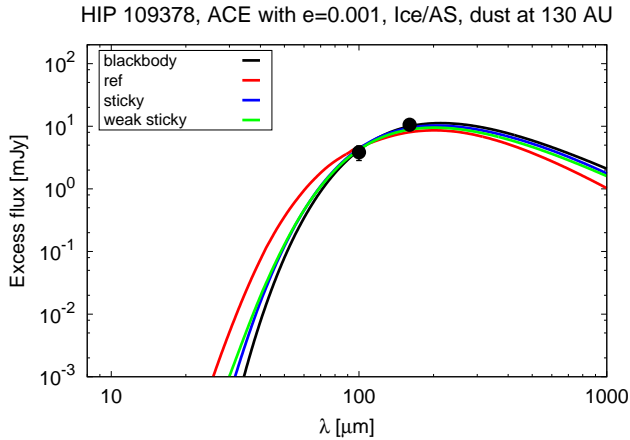}
  \end{center}
  \caption{
  SEDs of the dynamically cold belt of macroscopic grains
  (red, blue, and green lines correspond to those in Fig.~\ref{fig:model_size}).
  The ``blackbody'' SED (black line) and the observed excess emission of HIP~109378
  (symbols with error bars) are shown for comparison and 
  are the same as in Fig.~\ref{fig:sizes-materials}a,b.
  }
  \label{fig:model_sed}
\end{figure}

For the impact velocities and impactor sizes in question, the runs
show a rather complex mixture of outcomes, including disruption,
cratering and bouncing with mass transfer,
and agglomeration in comparable fractions.
The evolution of the size distribution over 5~Gyr for all three
runs is shown in Fig.~\ref{fig:model_size}. Bumps in the distributions at 
smaller sizes are numerical artifacts, coming largely from a limited resolution 
of the phase space grid.
These results demonstrate both the accretional growth of solids to sizes above $1\cm$
(particularly significant in the ``weak sticky'' run) and a moderate amount of 
collisional fragments in the sub-mm range. The latter, however, are still large
enough ($\sim 100\mum$) to stay sufficiently cold.
Indeed, the resulting SEDs, shown in Fig.~\ref{fig:model_sed}, are all
close to the ``blackbody SED''.
They are consistent at the 1-$\sigma$ level
with the available photometry points of HIP~109378.

We stress that the setup of these simulations is by far not unique and thus
the agreement with the data should not be overinterpreted.
The purpose of these simulations was solely to demonstrate
that the ``macroscopic belt'' scenario is conceivable and is potentially able to
explain the observations.

\subsection{Constraints on sizes}

We now try to paint a more general view.
Denote by $s_\mathrm{min}$ and $s_\mathrm{max}$ the ``effective'' minimum and the maximum radius of
the disk solids~-- i.e., the range of sizes beyond which the amount of particles
is too small to affect the dynamical evolution and thermal emission of the disk.
The largest objects have to be large enough to survive
against various loss processes over the systems' ages which, in case
of our cold disk candidates, are in the Gyr range.
One of these loss mechanisms is P-R drag (further mechanisms
are discussed in Sect.~5).
Assuming a unit radiation pressure efficiency,
the associated timescale is given by
\be
 T_\mathrm{PR} \sim 7 \left( L_\odot \over L_\star \right)
               \left( r \over 100\AU \right)
               \left( \rho \over 3\g\cm^{-3}   \right)
               \left( s \over 1\mm   \right) \Gyr ,
\ee
where $\rho$ is the bulk density of dust.
We conclude that $s_\mathrm{max} \ga 1\mm$ would satisfy the P-R drag-induced condition.

The lower limit on $s_\mathrm{min}$ from the ``coolness'' of the SED ($\ga 100\mum$)
and the lower limit on $s_\mathrm{max}$ from the P-R drag condition ($\ga 1\mm$)
are not the only constraints that we can place.
The upper limit on $s_\mathrm{max}$ can be estimated from
the requirement that typical random velocities of grains $v_{rel}$
do not exceed a few tens of $\m\s^{-1}$.
These random velocities cannot be lower than the relative velocities,
to which grains are stirred by the biggest bodies embedded in the disk.
The latter are approximately equal to the escape velocities from
the surface of the largest planetesimals with radius $s_\mathrm{max}$:
\be
  v_\mathrm{rel} > \sqrt{8/3 \pi G \rho} s_\mathrm{max} .
\label{eq:no stirring}
\ee
For the bulk density $\rho \sim 2 \g \cm^{-3}$, this leads to a simple result:
$s_\mathrm{max} < v_\mathrm{rel}$,
where 
$s_\mathrm{max}$ is in km and $v_\mathrm{rel}$ in $\m\s^{-1}$.
For instance, $v_\mathrm{rel}$ of $3\m\s^{-1}$ implies $s_\mathrm{max} < 3\km$.
For disks of a $\sim 100\AU$ radius around a solar-type star,
this sets the largest possible size of objects in the disk to
a few kilometers.

\begin{figure*}[hbt!]
  \begin{center}
  \includegraphics[width=0.77\textwidth]{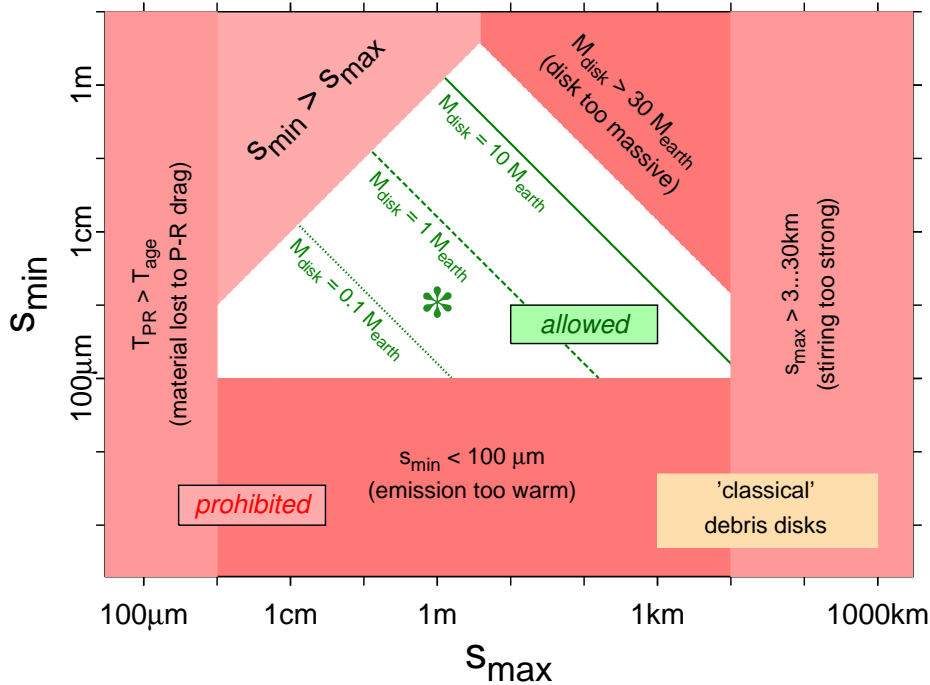}
  \end{center}
  \caption{
  Constraints on the minimum ($s_\mathrm{min}$) and the maximum ($s_\mathrm{max}$) size
  of objects in the cold disks in the proposed scenario of an ensemble of unstirred
  macroscopic grains.
  Unfilled region is allowed, filled one is prohibited.
  Lines of constant disk mass were calculated with
  Eq.~(\ref{eq:M_d}) for $f_d = 10^{-6}$.
  See text for details.
  The green asterisk corresponds roughly to the simulations described
  in Sect.~4.3.
  The domain of ``classical'' debris disks located in the bottom right corner
  is shown for comparison.
  }
  \label{fig:smin-smax}
\end{figure*}

Another requirement is that the total mass of the disk $M_d$,
which is also determined by the largest objects, should not become
unrealistically large. 
Assuming that the disk is composed of solids with sizes from $s_{min}$
to $s_\mathrm{max}$ with the size distribution $\propto s^{-q}$, where
$3<q<4$  and we take $q=3.5$ for simplicity,
it is easy to derive the relation between
$M_d$, the fractional luminosity $f_d$, as well as $s_\mathrm{min}$ and $s_\mathrm{max}$:
$M_d = {\rm const} \cdot f_d \cdot \sqrt{s_\mathrm{min} s_\mathrm{max}}$.
The prefactor can be directly estimated from the simulations described
in Sect.~4.3. For the HIP~109378 disk in the ``sticky'' run,
in which the size distribution can roughly be approximated
by a power law with $s_\mathrm{min} \approx 1\mm$,  $s_\mathrm{max} \approx 1\m$,
and $q \approx 3.5$, we obtained $f_d \approx 5 \times 10^{-6}$
with the disk mass of $M_d \approx 0.25 M_\oplus$.
This gives
\be
 M_d \approx 1.3 f_d \sqrt{s_\mathrm{min} s_\mathrm{max}} ,
\label{eq:M_d}
\ee
where $M_d$ is in the Earth masses and 
$s_\mathrm{min}$ and $s_\mathrm{max}$ in micrometers.
Total disk masses exceeding $\sim 10 M_\oplus$ would probably be unrealistic.
This is because the initial disk masses at protoplanetary stage
are typically comparable with the Minimum Mass Solar Nebula (MMSN) mass,
$0.01 M_\odot$, albeit with a large scatter
\citep[see][for a review of submillimeter observations]{williams-cieza-2011}
and, applying a standard 100:1 gas to mass ratio, are expected to contain
$10$--$100 M_\oplus$ of solids. Furthermore, the mass of the cold disks
is probably much less than that, because it only contains a fraction of
that material survived on the periphery of the systems.
For $s_\mathrm{min} \la 1\mm$, this implies a constraint on $s_\mathrm{max}$ nearly
as stringent as the ``no stirring''-constraint, Eq.~(\ref{eq:no stirring}):
$s_\mathrm{max} \la 10\km$.

Figure~\ref{fig:smin-smax} summarizes all these constraints, depicting
the ``allowed'' region in the $s_\mathrm{min}$--$s_\mathrm{max}$ plane.
It suggests that the observations are consistent with a hypothesis
of an unstirred belt of macroscopic solids, whose sizes may lie
somewhere in the millimeter- to kilometer-range.

\section{Conclusions and Discussion}

\subsection{Galaxies or disks?}

In this paper, we consider six ``cold debris disk'' candidates identified
by the {\it Herschel} OTKP DUNES. Since these observations
are at risk of being contaminated by chance alignments of extra-galactic sources,
we cannot definitely rule out a possibility
that any of these sources could be unrelated objects
rather than true circumstellar disks.
However, our catalog search for possible X-ray sources and their optical counterparts
in the fields around the optical positions of the six stars 
has not identified any sources that could be associated with contaminating galaxies.
Besides, a conservative estimate, based on the density of galaxies in the DUNES fields,
leads to the probability that some of the candidates are real disks
(and thus the ``cold debris disk'' phenomenon as such is real) of $\approx 99\%$.

Final answers can be found by repeating the observations of the candidates
with another instrument, ideally at more than one wavelength and in more than
one epoch, in order to reveal proper motion (or the absence thereof) of the sources
of emission.
For this purpose, one does not need to resolve the disk.
Such observations will certainly be possible, for instance with ALMA.
For point sources, $50\;\mu\mathrm{Jy}$ detections (one hour, $5\sigma$) at 
$1.3\mm$ (band~6) are possible,
and this would be more than sufficient to detect the cold disk candidates.
A successful ALMA detection after a few years 
might help to rule out background contamination 
via positional shifts of the source compared
to field objects (which we have from PACS/$160\mum$ or SPIRE/$250\mum$).
If there is still an object at the old ({\it Herschel}) stellar position,
it is likely a contaminating galaxy.
The only caveat to this is that ALMA's (sub-)mm sky will probably be quite
different to the {\it Herschel}'s far-IR sky~-- as it actually is
even among the three PACS bands.
Thus it would be better
to observe the fields with ALMA at two epochs, in order to approach the
problem under the same sky confusion background.

Beside  ALMA, detection with CCAT \citep{sebring-et-al-2006}\footnote{See also 
http://www.ccatobservatory.org.}
and {\it SPICA}/SAFARI \citep{goicoechea-nakagawa-2011}
should be possible.
The sensitivity (point-source, one hour integration, $5\sigma$)
of CCAT at its primary wavelengths of $350\mum$ and $450\mum$ should be about
$1\mJy$, and that of {\it SPICA}/SAFARI in the $100$--$200\mum$ range as high as
$30 \muJy$.
Besides, both instruments will be capable of performing a high-resolution spectroscopy.
This, especially in the case of {\it SPICA}/SAFARI with its excellent sensitivity, might
allow another method to establish the nature of cold emission~--
by detecting or ruling out characteristic spectral features that
might be typical of background galaxies.

Obviously, the most advantageous would be resolved observations of the candidates.
Due to the higher angular resolution, the probability of contamination in general would be lower.
The time span between two observations sufficient to confirm 
or rule out the common proper motion would be much shorter.
Last but not least, a detection of a ring-like structure around the star
(which might also be detected, depending on the setup and integration time)
would be undeniably the best proof that we are dealing with a disk. 

Which instrument has~-- or will have~-- a capability of sufficiently resolving the sources?
{\it SPICA}, with its 3.2 meter mirror, would have a resolving power
comparable to that of {\it Herschel}, and thus would not offer any advantages.
Using  CCAT could be more promising, as its 25 meter mirror would provide
a resolution of $2.9\arcsec$ and $3.7\arcsec$ at $350\mum$ and $450\mum$, respectively,
which is a factor of 3--4 better than that of PACS at $160\mum$.
However, the most promising instrument would be ALMA.
The expected total excess flux from the cold disk candidates is $\sim 1$~mJy at $1.3\mm$,
which would be reachable with a beam of $\approx 1\arcsec$.
Given the point-source sensitivity quoted above, the disk 
can be distributed over 20 beams to be detected.
Assuming an angular resolution of $1\arcsec$ and a face-on disk (worst case),
the dust can be distributed over a ring of $\la 3\arcsec$ in radius to be still detected 
at $5\sigma$. 
For the edge-on disk (best case), a radius of $\la 10\arcsec$ would suffice.
The range $3$--$10\arcsec$ is approximately what we expect for the cold disks from
the {\it Herschel} data.

\subsection{If disks, are they dynamically cold?}

For true circumstellar disks, and assuming that the gravitational perturbers
(planets, substellar or stellar companions) are absent, the proposed scenario of
dynamically cold (i.e., unstirred) disks would imply a rotationally symmetric,
ring-like confinement of material and a lack of any observable offsets or clumps.
If a perturber is present, the disk could still be dynamically cold if
the disk particles shared the same {\em forced} eccentricities induced by the
perturber, but had {\em proper} eccentricities close to zero,
keeping the relative velocities at a very low level.
It may be possible if the particles' eccentricities were initially
close to the forced eccentricity from the perturber, and remained there for the
disk's lifetime. This may be achievable if the particles' eccentricities were
damped to the forced eccentricity when a gas disk was still present
\citep{thebault-2012}, or if collisions between particles occur frequently enough
and at low enough velocity to damp eccentricities in the collisions without
destroying the particles.
The disk would be elliptic, similar to the Fomalhaut ring
\citep{kalas-et-al-2008,chiang-et-al-2009}.
In this case, there would be an offset, and possibly a pericenter glow
\citep{wyatt-et-al-1999}.

Verifying these possibilities appears to be a more difficult task than confirming
the circumstellar origin of the emission, because it requires the disk
to be well-resolved.
However, as discussed in Sect. 5.1, sufficiently well resolved observations
with ALMA seem possible.

\subsection{Possible implications for planetesimal formation?}

Assuming that the observed emission is indeed 
related to the stars, we concluded that the emitting material should have nearly a blackbody temperature.
We then examined possibilities to explain
why the emitting material in these disks is thermally cold.
We argued that the cold disks should be composed of weakly stirred or unstirred
primordial solids with radii in the range from millimeters to about
ten kilometers. Tighter constraints on sizes are difficult to pose,
but solids larger than $\sim 1\mm$ are needed for the material to survive 
against P-R drag,
while objects larger than $\sim 10\km$ must be absent, since these would
stir the disk out of its ``cold'' state.
Nor is it possible to put more stringent constraints on the degree of
dynamical excitation in the disks.
We showed that the maximum eccentricities and inclinations should not exceed $\sim 0.01$,
since the amount of small dusty debris would otherwise contradict
the observational data.
However, these might easily be as low as $\sim 0.001$, corresponding to
relative velocities of a few $\m\s^{-1}$.
Our simulations suggest that a system with $e \sim I \sim 0.001$--$0.01$ would
experience gentle collisions that only involve a moderate production of small
dusty debris and a moderate amount of (further) accretional growth, and could
survive around a star for gigayears. 
A principal possibility for systems of this type to exist
was recently pointed out by \citet{heng-tremaine-2010}, and Saturn's rings  
in the solar system readily provide an example of such
disks, albeit on a much smaller spatial scale \citep{esposito-2002}.

We now compare our findings with models of planetesimal growth.
At early phases, all these models have to overcome various hurdles.
One is a rapid loss of material due to radial drift at some sizes
\citep[e.g.,][]{weidenschilling-1980} and another
is a switch from agglomerational to fragmentational regime at growing
sizes \citep[e.g.][and references therein]{blum-wurm-2008}.
Under the MMSN conditions at $\sim 1\AU$, both are expected to happen
at sizes of about one meter and thus are often referred to as 
the ``meter-barrier''.
However, in the outer parts of the systems (at $\sim 100\AU$),
drift is the fastest at millimeters or centimeters
rather than decimeters or meters \citep{brauer-et-al-2007}.
Various ways have been suggested to circumvent both the drift and fragmentation
hurdles.
The drift problem may not exist in turbulent disks,
and models with more and more realistic physics involved may also help
to eliminate or at least mitigate the fragmentation problem
\citep[e.g.][]{brauer-et-al-2008,zsom-et-al-2010,zsom-et-al-2011}.
Alternative pathways for a rapid unimpeded growth have also been proposed that
invoke collective dust phenomena, such as capture of solids in pressure maxima of
the gas disk \citep{johansen-et-al-2006}, possibly enhanced by
streaming instability \citep{johansen-et-al-2007}, both with subsequent
local gravitational clumping of material
\citep[see][for a review]{chiang-youdin-2010}.
Whatever path the system takes~-- if kilometer
or larger sizes have been reached, further growth of planetesimals is expected in 
gravity-assisted pairwise collisions \citep[e.g.][]{goldreich-et-al-2004}.

All models naturally predict the planetesimal growth to strongly depend on the
distance from the star.
At any given age, the largest objects are smaller farther out and conversely,
it takes longer for the planetesimals to grow to a given size farther out from the
central star. Effectively, the growth should stall at large distances from the star,
where the nebula has a very low density.
Furthermore, protoplanetary disks have finite sizes, being truncated by internal
physical processes or external influences such as stellar encounters.
These sizes are probed by observations
\citep[see][for a review]{williams-cieza-2011}.
(Sub)millimeter data suggest exponentially tapered outer edges of disks to lie
between a few tens and a few hundreds of AU, with radii of $\sim 100\AU$ being
typical \citep[e.g.][]{andrews-et-al-2009,andrews-et-al-2010}. A similar range
of radii, $\sim 50$ to $\sim 200\AU$, comes from direct measurements of silhouette
proplyds \citep{vicente-alves-2005}.
With their estimated radii of $\sim 50$--$150\AU$, the cold disks
may thus simply trace the outer edges of the original planetesimal disks, 
namely the maximum distance up to which planetesimals could form.

More specific conclusions are extremely difficult to arrive at.
As explained above, we cannot predict whether solids in the cold disks
are millimeters or kilometers in sizes. We cannot even be sure whether they entered
the gravity-driven growth regime.
Previous observations of protoplanetary disks do tell us that dust growth
on the periphery of the systems advances at least to millimeters 
\citep[e.g.][]{wilner-et-al-2005,rodmann-et-al-2006},
but they do not probe larger sizes.
The model predictions for the planetesimal formation timescales and final sizes
are extremely uncertain, too.
\citet{kenyon-bromley-2008}, for instance, assumed initial planetesimal sizes
of $1\m$--$1\km$ and modeled planetesimal growth in the
$30$--$150\AU$ range around $1$--$3M_\odot$ stars in a set of possible nebulae
with $(0.3$--$3)\times$MMSN density profiles.
They found that after $\sim 1\Gyr$ of evolution the largest planetesimals in
an MMSN disk reach sizes of $\sim 100\km$ even at $r\sim 150\AU$.
This might conflict with our conclusion that the planetesimal growth in cold disks
must have stopped before these ``cometary'' or ``asteroidal'' sizes have been reached.
The possible controversy can, however, easily be mitigated by varying the model
assumptions, for instance the density profile of the original disk. Assuming
$0.3\times$MMSN reduces the maximum sizes to $\sim 30\km$, close to what is needed.
Alternative models of rapid planetesimal formation that invoke clumping of material
in turbulent disks, are not certain in their predictions either.
For instance, \citet{johansen-et-al-2012} find the final sizes of Kuiper objects
formed in this way to be $150$--$730\km$,
but it is easy to imagine that the same mechanisms at distances of $\sim 100\AU$
and/or under different model assumptions could halt at smaller sizes.

\subsection{Planets in the cold disk systems?}

Cold disks must have large inner voids, as inferred from the absence of emission
at $\la 100\mum$.
In the inner parts of the ``cold disk'' systems,
the planetesimal formation is likely to have advanced further,
ending up with formation of planets (Marshall et al., in prep.).
Indeed, one of the cold disks, HIP~109378, was reported to host a radial velocity 
planet \citep{marcy-et-al-1999}.

Whether planets are present farther out from the stars in the inner gaps
of the cold disks is unclear.
If they were, they would naturally account for these inner voids.
However, as discussed in Sect. 5.2, the presence of planets would have to be
compromised with the suggested dynamically cold state of the disks by requiring low proper
eccentricities of the disk particles from their formation stage or
from the subsequent collisional damping.
And conversely, if the planets were absent, the disks could preserve their 
dynamically cold state easily.
However, this would raise
the question of what, if not planets, has created the inner voids.
A few mechanisms that do not necessarily involve planets appear feasible.
These may include for instance dust drag triggered by UV-switch 
\citep[e.g.][]{alexander-armitage-2007}
and collisional depletion of the inner region of an initially extended
planetesimal disk \citep[e.g.][]{wyatt-et-al-2012}.

Interestingly, one of the six stars, HIP 92043, shows a K-band excess of around 1\%,
which has been interpreted as stemming from
a hot exozodiacal cloud \citep{absil-et-al-2013}.
Such exozodis have previously been detected
with CHARA/FLUOR around Vega, Fomalhaut, $\tau$~Cet and some other stars
\citep[see][for a recent review]{absil-mawet-2010}.
The origin of these exozodis is as yet unclear; these may or may not
be related to the outer disks \citep{bonsor-et-al-2013}.

\subsection{Could disks survive and remain cold for gigayears?}

Most of the above discussion is about the very early stages of planetesimal formation.
From solar system \citep{morbidelli-2010} and exoplanet system studies
\citep{booth-et-al-2009,raymond-et-al-2011,raymond-et-al-2012} we know that a 
planetary system may undergo
violent dynamical rearrangements that might smear out or substantially alter
the architecture left immediately after the completion of the protoplanetary
phase. Even if the systems evolve in a smooth way, cold belts on the periphery
might be threatened by a variety of effects over Gyr timescales. 
These include possible interactions of disks with the ISM, especially
during passages through clouds, and galactic tidal forces.

For HIP~171, which is a binary star, the cold disk~-- if it is real~-- would
be circumbinary. Since the binary has a substantial eccentricity $(0.38)$, it
is not clear whether the outer disk could have remained unstirred over the system's age,
even though the semimajor axis ratio is large \citep{mustill-wyatt-2009}.
Simulations for a disk of particles with a radius of $R_\mathrm{disk}= 60\AU$
(Tab.~\ref{tab:six_data})
initially in orbits with eccentricities between 0 and 0.1 showed that
after 1 Gyr the eccentricity dispersion of the particles, induced by the binary,
is of the order of 0.03--0.04 (Mustill et al., in prep.).
This is larger than what is required by the dynamically cold scenario.
Still, there exists a possibility to keep the eccentricity oscillations low
if the particles' eccentricities are initially close to the forced
eccentricity from the binary, and remain there for the disk's
lifetime~-- see a discussion in Sect. 5.2.

\subsection{Concluding remarks}

We conclude by noting that,
regardless of whether all of the ``cold disks'' are true disks or some are
unrelated background sources,
the theoretical analysis of collisional evolution in different dynamical
regimes presented in Sect.~4 of this paper may be of generic interest.
This is because we try to find out how 
debris disks could operate at very low optical depths and at
a low level of stirring, as can particularly be expected for disks
with large radii.
Both domains comfortably lie in the discovery
space of the far-IR, sub-mm, and radio facilities that are starting to
operate (such as SCUBA-2 and ALMA) or are being developed (e.g. {\it SPICA}).
Thus our analysis might provide useful guidelines for interpretation of
the data that are expected to come.

\section{Summary}

\begin{itemize}
\item
In this paper, we consider six ``cold debris disk'' candidates identified
by the {\it Herschel} OTKP DUNES
and argue that, at a high level of confidence, most of these candidates
represent true circumstellar disks.
\item
For true circumstellar disks, the available data suggest
that the dominant size of the grain cross section is larger than $\sim 100\mum$
and that the smaller grains are strongly underabundant.
This contrasts with all of the debris disks observed previously,
where observations and models both reveal dominant sizes to lie
in the micrometer range.
\item
A plausible explanation for the dearth of small grains is the unstirred
disks of solids grown on the periphery of systems during the protoplanetary phase.
We show that, to explain the data,
they should comprise solids larger than millimeters, but smaller
than a few kilometers in size.
This would imply that planetesimal accretion, at the
least in outer regions of the systems, has stopped before ``cometary''
or ``asteroidal'' sizes were reached.
\end{itemize}

\bigskip
{\small
{\bf Acknowledgments:}
We are grateful to the referee for his/her useful comments that greatly
helped to improve the manuscript.
A.V.K. and T.L. thank J\"urgen Blum and Carsten G\"uttler for their explanations
on collisional outcomes and Ludwig Trepl for discussions on the {\it ROSAT} data.
Work of A.V.K., T.L., and S.W. was partly funded by the {\em Deutsche
Forschungsgemeinschaft} (grants Kr 2164/10-1, Lo 1715/1-1, and Wo 857/7-1). 
C. E., J. P. M., and B. M. were partly supported by Spanish grant 
AYA 2011-26202.
J.-C.A. and S.E. acknowledge financial support of the CNES-PNP.
S.E. also thanks the French National Research Agency (ANR) for financial support
through contract ANR-2010 BLAN-0505-01 (EXOZODI).
A.B. was co-funded under the Marie Curie Actions of the European Commission
(FP7-COFUND).
}


\newcommand{\AAp}      {A\&A}
\newcommand{\AApR}     {A\&AR}
\newcommand{\AApS}     {A\&AS}
\newcommand{\AApSS}    {A\&AS}
\newcommand{\AdvSR}    {AdvSR}
\newcommand{\AJ}       {AJ}
\newcommand{\AN}       {AN}
\newcommand{\AO}       {AO}
\newcommand{\ApJ}      {ApJ}
\newcommand{\ApJL}     {ApJL}
\newcommand{\ApJS}     {ApJS}
\newcommand{\ApSS}     {Ap\&SS}
\newcommand{\ARAA}     {ARA\&A}
\newcommand{\ARevEPS}  {ARE\&PS}
\newcommand{\BAAS}     {BAAS}
\newcommand{\CelMech}  {CMDA}
\newcommand{\EMP}      {EMP}
\newcommand{\EPS}      {EPS}
\newcommand{\GRL}      {GRL}
\newcommand{\JGR}      {JGR}
\newcommand{\MemSAI}   {Mem. Soc. Astron. Ital.}
\newcommand{\MNRAS}    {MNRAS}
\newcommand{\PASJ}     {PASJ}
\newcommand{\PASP}     {PASP}
\newcommand{\PSS}      {PSS}
\newcommand{\RAA}      {RA\&A}
\newcommand{\SSR}      {SSR}



\end{document}